\documentstyle[aps,pre,epsfig,twocolumn]{revtex}   
  
\begin{document}    
   
\wideabs{   
    
\title{Granular Shear Flow Dynamics and Forces:    
Experiment and Continuum Theory}    
    
\author{L. Bocquet$^{1,3}$, W. Losert$^{2,}$\thanks{Current address:    
Department of Physics, University of Maryland, College Park, Maryland 20742},   
D. Schalk$^{2}$, T.C. Lubensky$^{3}$ and J.P. Gollub$^{2,3}$}    
   
\address{$^{1}$ Laboratoire de Physique de l'E.N.S. de Lyon, U.M.R.    
C.N.R.S. 5672, 46 All\'ee    
d'Italie, 69364 Lyon Cedex, France}   
\address{$^{2}$ Physics Department, Haverford College, Haverford, PA 19041}   
\address{$^{3}$ Physics Department, University of Pennsylvania,   
Philadelphia, PA 19104}   
  
\date{\today}      
\maketitle   
\begin{abstract}   
We analyze the main features of granular shear flow through experimental   
measurements in a Couette geometry and a comparison to a   
locally Newtonian, continuum model of granular flow.     
The model is based on earlier hydrodynamic models, adjusted to take    
into account the experimentally observed    
coupling between fluctuations in particle motion    
and mean-flow properties.  Experimentally, the local velocity    
fluctuations are found to vary as a power of the local velocity    
gradient.    
This can be explained by an effective viscosity that diverges more rapidly   
as the random-close-packing density is approached than is predicted by  
Enskog theory for dense hard sphere systems.   
Experiment and theory are in good agreement, especially    
for the following key features of granular flow: The flow is confined to a    
small shear band, fluctuations decay approximately exponentially away from    
the sheared wall, and the shear stress is approximately    
independent of shear rate.   
The functional forms of the velocity and fluctuation profiles predicted   
by the model agree with the experimental results.   
\end{abstract}   
    
}   
   
\section{Introduction}   
   
The general features of granular shear flow have been investigated thoroughly    
over the last several years~\cite{clement99,Behringer99}.     
The following key points emerge in shear flow experiments    
on a large range of materials in two and three dimensions:   
   
\begin{itemize}   
\item The velocity of particles decreases quickly over a few particle    
diameters away from the shearing    
wall (see e.g. ~\cite{Behringer99,Mueth2000,Losert2000}).   
\item The velocity profile, normalized by the shear rate $U$,    
is independent of $U$ (see e.g. ~\cite{Behringer99,Mueth2000}).   
\item The shear force $\sigma$ is approximately independent of $U$,    
if the granular material is allowed to dilate (see e.g.~\cite{Tardos98}).   
\end{itemize}   
   
These features, together with the discovery of strong inhomogeneities in   
the force distribution even during flow~\cite{Behringer99,Aharonov99},     
might be taken to indicate that any continuum approach, such as local     
hydrodynamic models, should fail to describe granular flow.   
   
Here we revisit the assumptions made in earlier hydrodynamic models of    
granular flow, via careful comparison to the experimentally measured    
microscopic particle dynamics in a circular Couette geometry.  This leads us to emphasize the strong    
interplay between local RMS fluctuations, the mean flow, and    
the local density.    
When this coupling is properly taken into account, a hydrodynamic   
model, which we have introduced in Ref.~\cite{ LosertPRL},   
quantitatively describes all key properties of granular shear flow   
discussed above, including both flow properties and shear forces.   
   
The shear force obtained from the hydrodynamic model    
resembles the simple dynamic friction law found in solid-on-solid friction,    
i.e. the shear force is proportional to pressure, but approximately    
independent of   
shear rate.  We emphasize that this result is obtained even though the    
hydrodynamic model does not include frictional forces between grains.    
   
The temperature is defined here as the mass times the   
square of RMS velocity fluctuations.     
While the temperature is a constant in shear flow of an ordinary fluid,   
granular temperature is dissipated through inelastic collisions.  Its    
spatial variation plays a crucial role in determining the properties of  
granular flow.     
We find that the granular temperature profile, normalized by its maximum    
value, is independent of shear rate and pressure.     
Temperature is introduced into the system via viscous heating  
over a characteristic length    
of the order of a few particle diameters.      
It is then dissipated via inelastic collisions over a longer length  
scale.  
Our model contains a description of the source and transport   
of fluctuations that allows us to predict the pressure and    
the shear rate dependence   
of both the shear forces and the particle dynamics.    
   
In the experiments reported here, which go beyond those reported    
earlier~\cite{ LosertPRL}, the granular material is sheared in a Couette   
geometry with a rotating inner cylinder and a stationary outer cylinder.     
The inner cylinder is connected to the motor   
through a flexible spring that allows either stick-slip motion or   
steady shearing depending on parameters.   
We can also apply an upward air flow at a variable rate through the granular   
material to dilate the material and reduce the stresses.   
Our ability to vary the stresses via air flow, and our study of both   
stick-slip and steady dynamics in the same apparatus,  
distinguish this work from other    
experimental measurements of sheared    
granular matter~\cite{Mueth2000,Behringer99}.    
   
In addition to performing force measurements, which probe macroscopic  
material properties of the ensemble of particles,   
we determine the dynamics of individual particles by measuring   
the mean velocity and RMS velocity fluctuations of particles on the top    
surface of the granular layer.   
The combination of velocity and force measurements, together    
with variation of the stresses and the time dependence of the flow,   
allow for a very sensitive test of our theoretical model.   
   
Previous experimental results and modeling approaches are reviewed in   
section~\ref{backgroundchapter}. Our experimental setup and results   
for the particle dynamics and shear forces in sheared granular matter are   
discussed in sections~\ref{expreschapter} and~\ref{expsec2}.     
In section~\ref{theorychapter},   
the locally Newtonian hydrodynamic model is described in detail.   
We conclude in section~\ref{discussionchapter} with a discussion   
of the main results and the broader implications of this work.   
   
\section{Background}     
\label{backgroundchapter}     
     
Efforts to understand particle dynamics during granular flow      
can be roughly divided into adaptations of continuum models,     
based on hydrodynamic or elasto-plastic descriptions, and     
models that emphasize the differences between molecular fluids or solids     
and granular matter, such as the inhomogeneous character of particle      
contacts and of stress transmission in a granular material.     
In this section we review briefly these different approaches.    
We conclude with a discussion of the aim of our hydrodynamic    
model in the context of previous work.    
     
\subsection{Continuum models of granular flow}     
   
   
Hydrodynamic models were motivated by Bagnolds pioneering theoretical     
and experimental work on shear forces in    
dense suspensions~\cite{Bagnold1954}.   
The constitutive equations for macroscopic quantities,      
such as the shear stress as a function of shear rate, were   
investigated in several studies of dense suspensions with fixed volume.    
In the limit of large velocities,    
Bagnold found that in a fixed volume the shear stress $\sigma$      
is proportional to the shear rate $U$ squared.   
He referred to this regime as the ``grain inertia regime''.  
He accounted for the measurements     
by assuming that the local shear stress $\sigma$ was proportional to    
the square of the local shear rate $\dot\gamma$.     
He justified such a relationship on the basis of    
kinetic arguments.    
Since direct measurements of microscopic particle dynamics were not available,     
he assumed a linear velocity profile similar to that of ordinary fluids.   
   
The development of hydrodynamic descriptions, extracted from the   
``microscopic'' dynamics using kinetic   
theory, was pioneered by Jenkins and Savage~\cite{jenkins83} and  
by Haft~\cite{haft83}.    
A considerable amount of theoretical, numerical, and experimental   
work has refined and modified this approach.   
Reviews of models of granular flow that describe many   
of these studies has been compiled by Campbell~\cite{Campbell90},   
Savage~\cite{Savage1993}, and more recently by Clement~\cite{clement99}.   
   
In the kinetic theory approach,   
the granular material is generally modeled as an inelastic hard-sphere   
system.  Constitutive equations similar to the usual Navier-Stokes equations   
of hydrodynamics can be obtained in the limit of small    
inelasticity.  The transport coefficients entering the flow equations   
are usually computed at the level of the Enskog equation \cite{JPH}, 
an extension of the Boltzmann equation that takes   
the finite size of particles into account but neglects correlations    
between collisions.   
Due to the assumptions that enter into the kinetic theories presented   
above, these descriptions are limited to rapid granular flows and   
to intermediate or low particle densities.     
       
In order to understand the boundary between a seemingly flowing state and     
an apparently stationary state,  
various extensions of the previous hydrodynamic model have been proposed. 
Jenkins and Askari \cite{Jenkins90} have studied the interface between a 
flowing region and an amorphous (high density) region which is at rest. 
In this model, the thickness of the shear band 
is determined by the balance 
between the energy input and the loss rate due to inelasticity. 
On the other hand, 
visco-plastic models have been proposed.    
The Savage-Hutter model~\cite{Savage1989}   
uses the Mohr-Coulomb failure criterion    
to predict the transition from solid-like to fluid-like behavior in the   
context of avalanching and rock slides.   
The constitutive relation connecting shear stress to shear rate is of  
great practical importance.  
Different relations have been proposed for various situations, and   
several are summarized in a table in~\cite{Hutter1993}.     
   
\subsection{Alternative descriptions of granular flow}     
    
Beyond measurements of the mean properties (such as shear forces) that 
are important for a continuum model,     
recent experimental and theoretical studies     
have focused on measurements of the particle dynamics and    
shear forces at the scale of an individual particle.      
    
Detailed measurements of the particle     
dynamics (see e.g.~\cite{Behringer99,Mueth2000,Losert2000})     
revealed that in several experimental geometries particle motion is confined    
to several (generally 5-10) particle diameters close to the sheared surface.    
The velocity profile is found to be roughly exponential or Gaussian.  
    
The force distribution within a granular assembly, measured with birefringent    
disks~\cite{Behringer96} or carbon paper~\cite{Nagel98}, was found to exhibit    
strong inhomogeneities on the particle scale.  Stresses were found to be    
transmitted along chains of particles (force chains) in a static    
granular assembly and during shear.      
   
{There have been several attempts} 
to account specifically for these inhomogeneities    
in granular flows.  Some approaches describe the flow properties on the    
basis of fracture models~\cite{pouliquen,josserand}, while others    
introduced non-local constitutive equations coupling force chains to    
flowing grains~\cite{Mills99}.     
   
The shear strength of deformable, inelastic spheres was modeled using     
a discrete element method by Aharonov and Sparks~\cite{Aharonov99}.      
The density is found to adjust to a critical density within     
the shear band.      
     
\subsection{Our hydrodynamic model}     
   
While theories and models of granular flow have  {become remarkably detailed},  
the assumptions of each model can strongly influence    
the results.  As noted by Campbell~\cite{Campbell90},   
detailed experimental measurements of particle velocities,   
granular temperatures and densities were often unavailable   
when models were developed.  The measurements reported here  
should help assess the validity of the assumptions of various models.    
   
Here we revisit the local hydrodynamic model and carefully reexamine the     
assumptions made to derive the constitutive equations in view of our      
experimental measurements of individual particle dynamics and mean    
shear forces.     
    
The mean shear forces have been studied before in an experimental    
system similar to ours:  Tardos {\it et al.}~\cite{Tardos98}   
investigated the effect of an upward air flow through the granular material    
on the shear forces.  The shear force was found to decrease linearly    
with air flow.     
On the basis of Bagnold's results~\cite{Bagnold1954} and much subsequent      
work, it is clear that dilatancy has an important     
effect on granular flow.  This was also convincingly demonstrated   
in the experiments by Tardos {\it et al.}:   
If the material is allowed to expand,  
the shear stress is independent of shear rate.   
On the other hand, if the material is confined to a fixed    
volume, Bagnold's result of a quadratic increase of shear stress with    
shear rate is found.     
   
The interplay between dilatancy, fluctuations, and flow is complex.  
Theoretical investigations have not yet produced results  
that are consistent with measured mean (macroscopic) properties and  
measured dynamics at the particle (microscopic) scale.   
Granular material develops a greater resistance to flow as its density  
increases.  Previous theoretical treatments of granular flow have either  
been restricted to the lower-density  
rapid-flow regime \cite{clement99}, or they have  
incorporated a yield threshold (visco-plastic models) that produces a  
well-defined transition from liquid-like to solid-like behavior \cite{Savage1989}.  
   
Recent experimental studies of granular flow down a sandpile     
by Nasuno {\it et al.}~\cite{Nasuno2000} using long exposure time video    
imaging have revealed that the transition between solid-like and    
fluid-like behavior may not be very well defined.     
The velocity profile     
within the flowing layer was found to be exponential over more than seven   
orders of magnitude in speed with no clear transition to a solid-like state.     
These results indicate that no strict transition may exist, but     
instead one may be able to treat the solid-like state as a very    
high viscosity fluid.    
    
In our model we do not distinguish between a solid-like and fluid-like   
state. Rather, we try to model increased resistance to flow in the  
high-density   
limit within the standard kinetic approach (i.e., we assume   
binary collisions, no friction, etc.).   Previous  
approaches \cite{jenkins83,haft83}   
used the Enskog equations to take excluded volume    
at higher densities into account.  However, correlated   
motion of particles, e.g. through cooperative rearrangements, is not    
taken into account in the Enskog equation.   
   
The high density regime of polydisperse spheres is bounded by   
the largest possible random close packing (RCP), which experimentally   
is found to be $63.7\%$ for slightly polydisperse systems~\cite{Scott69}.   
For a wider distribution of particle sizes, slightly larger    
densities can be reached.   
In order to model the high density regime close to RCP, we use the    
results from elastic systems as a first approximation.     
In elastic systems, simulations in Lennard-Jones   
systems~\cite{Glotzer98} and experiments on colloids~\cite{Weeks2000}   
suggest that cooperative rearrangements are important at high density.    
In these systems the viscosity is found to    
diverge anomalously strongly with density in the high density limit 
~\cite{Alder1970,Leutheusser1982}.   
The divergence with density is roughly equivalent to    
the divergence with decreasing temperature for a supercooled   
liquid close to the glass transition~\cite{Weeks2000}.   
   
The introduction of a viscosity that diverges more rapidly than the thermal  
conductivity and heat loss coefficient as  
random close packing is approached is the key new feature of our model.  
It leads to theoretical predictions in quantitative agreement with most of  
our experimental results.    
It also provides a direct connection   
between the dynamics of granular media and glasses, as   
has been proposed by Liu and Nagel~\cite{LiuNagel}.   
   
\section{Experimental Setup}   
 \label{expreschapter}   
   
\subsection{Apparatus}   
   
In the experiments we shear the granular material in a Couette geometry.   
The granular material used in most of the experiments    
reported here consists of $0.55-0.95  {\rm~ mm}$    
diameter black glass beads (from Jaygo Inc.) ($\rho_m = 2.55 {\rm~ g/mm^3}$).     
The color does not alter surface properties, but increases the opacity of   
the material, which facilitates the tracking of particles on the surface    
as described below.  We also carried out experiments    
with a mixture of $1.3 {\rm~ mm}$ and  $1.6 {\rm~ mm}$ Chrome Steel spheres    
($\rho_m = 5.0 {\rm g/mm^3}$),   
and with polydisperse, rough ceramic spheres    
(Macrolite ML1430 from Kinetico Corp.)   
with diameter $0.83 -1.47 {\rm~ mm}$ ($\rho_m = 0.51 {\rm g/mm^3}$).     
      
\begin{figure}[t]   
 \begin{center}   
\epsfig{file=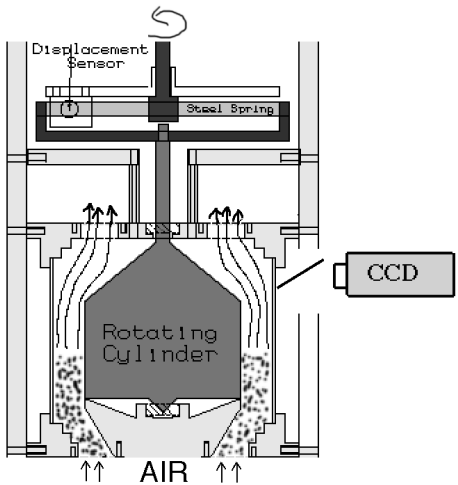,width=8.5cm}   
\end{center}   
\caption{Experimental setup:  The granular material    
(between two concentric cylinders)   
is fluidized by an upward air flow and sheared by rotation of the inner    
cylinder, which is connected to the motor through a flexible spring.    
Shear forces are determined from the spring    
displacement.  Particle motions in the top layer are measured   
through the glass outer cylinder with a fast CCD camera.   
 }   
\label{exp_setup}   
\end{figure}

In the experimental apparatus the granular material   
is confined to a $12 {\rm~ mm}$ gap between    
a stationary outer cylinder and a rotating inner cylinder ($r=51 {\rm~ mm}$),    
as shown in Fig.~\ref{exp_setup}. The gap can be reduced to $3 {\rm~ mm}$.   
The inner cylinder is hollow to reduce its inertia and is   
coated with a monolayer of randomly packed glass beads    
to provide a rough boundary.     
The outer glass cylinder is coated with a     
monolayer of randomly packed glass beads up to the    
height of the top surface,   
which allows observation of the top layer of grains through a mirror as   
shown in Fig.~\ref{exp_setup}.     
The lower $38 {\rm~ mm}$ of the inner cylinder is stationary    
in order to minimize boundary layer effects.   
   
To shear the material, the inner cylinder is rotated with a $4000$   
step/turn microstepping motor (from Aerotech Inc.)   
at a variable rate of $0.001 - 1$~Hz. The rotation rate is   
smoothed by a $100:1$ gearhead for rates $< 1$~Hz.   
The inner cylinder is connected to the microstepping motor    
via a flexible tempered steel spring.   
This spring configuration allows us    
to measure instantaneous shear forces with excellent dynamic    
range and precision, since the spring bending is proportional   
to the applied shear force.  We measure the spring bending    
with a capacitive displacement sensor     
(EMD1051, Electro Corp.) that is rigidly    
connected to the motor shaft at a radial distance of $4.2$~mm from the shaft.   
The spring constant of    
the spring (dimensions: $0.51 \times 7.5 \times 165$~mm)   
was determined to be $220 \pm 8 $~N/m.    
   
The soft connection between    
the motor and the inner cylinder   
permits both stick-slip dynamics and continuous motion of the    
inner cylinder to be obtained,    
depending on parameters. However, when      
a uniform speed of the cylinder is required for the experiment,    
the spring is replaced with a rigid connection. In that case   
measurement of shear forces is not possible.   
   
We can also apply an upward air flow at a variable rate through the granular   
material. The air flow enters the granular material through   
a circular opening between the cylinders, and leaves the granular material   
 through a circularly symmetric opening at the top.     
This assures a uniform air   
flow rate throughout the material.  Flow rate uniformity    
was tested by observing  the position of air bubbles   
as they leave the upper surface of the granular material.   
Air bubbles form at high flow rates in the class of granular materials used in    
this experiment.     
The random position of air bubbles indicates that the    
flow is uniformly distributed throughout the gap.   
Except for this test for flow uniformity,   
none of the experiments presented in this paper was carried    
in the presence of air bubbling.  
The air flow is provided by a blower (Rigid Inc.)    
operated at variable input AC voltage.  Flow rates are measured    
by means of    
an air velocity transducer (FMA 904, Omega Inc.).  For the   
range of air flow rates we employed, the average density   
of the granular material  
changes by less than $10$ \%  
and the flow speed is calculated in first approximation 
assuming the porosity for random close packing.   
   
Since the air flow exerts a drag on individual grains, the effective   
weight supported by neighboring grains decreases with increasing upward   
airflow.  This effective weight of individual    
grains in turn should be proportional to the pressure inside the    
granular material.    
We can, therefore, reduce the pressure by applying an upward air flow,    
and increase pressure by applying   
a downward air flow.  The proportionality   
factor between air flow and pressure can be roughly estimated   
by calculating the upward drag exerted on a single sphere at the    
mean air flow speed within the granular material~\cite{neglect}.    
   
\subsection{Determination of particle dynamics}   
   
We measure the mean particle velocities $V(y)$ and   
the velocity fluctuations $\delta V(y)$ on the upper surface   
of the granular material.  These should approximate    
particle motion in the interior based on a previous    
measurements~\cite{Mueth2000} that   
found very similar velocity profiles in the interior    
(measured with MRI and X-ray techniques) and on the bottom    
surface of a shear cell. We return to this issue in Sec. VI.  
   
The trajectories of    
individual particles in the surface layer are determined with a fast    
CCD camera at $30-1000$~frames/sec.  Particle motion is extracted    
from four sequences of $2000$ images   
 using procedures written in IDL (RSI Inc.) based on tracking    
routines provided by J. Crocker and E. Weeks.     
   
In the first step of the tracking process,    
both long range brightness fluctuations (i.e. nonuniform illumination)  
and short wavelength   
noise are reduced by applying a bandpass filter with a short wavelength   
noise cutoff and a long wavelength cutoff   
of roughly one particle diameter.   
In the second step, the position of particles (roughly $100$ in each image)    
is determined   
by calculating the centroid of each bright region in the filtered    
image.  This yields a spatial resolution of $< 0.1$ pixels, provided that the   
bright region is several particle diameters wide.   
Black glass beads are better suited than undyed glass beads   
for an accurate determination of   
particle positions, since black beads are more opaque. This   
reduces internal reflections and reflections from particles in deeper    
layers.  In order to improve spatial resolution,    
the intensity peaks are broadened by taking images slightly out of focus.   
The broader intensity peaks improve the precision of the    
centroid determination.  Defocusing also reduces the intensity of    
secondary peaks due to scattering by multiple particles to a level where 
they are no longer interpreted as particles.   
For the ceramic particles the defocusing process eliminates multiple    
peaks due to the substructure of individual particles.     
   
In the third step of the trajectory determination,    
the particles are labelled and the evolution of their position    
through an image sequence is determined.  The assignment of particles to   
corresponding points in the previous and next frame is   
based on a tracking algorithm, which    
minimizes the total squared displacement within a sequence of frames.     
In a final step, the probability distribution   
of individual particle displacements is used to verify that   
large displacement particles are not systematically cut off.

Since the mean-flow velocity $V$ is comparable to the RMS velocity   
fluctuations close to the inner cylinder, accurate tracking of particles   
is only possible if the maximum displacement is considerably smaller than the   
distance between particles.  We have verified that particles are accurately   
tracked, even when the maximum particle displacement between frames    
approaches the particle spacing.   
      
>From the particle tracks we determine average particle velocities $V(y)$ and    
RMS velocity fluctuations perpendicular to the flow direction   
 ($\delta V_x(y)$) and parallel to it ($\delta V_y(y)$)   
as a function of distance $y$ from the rotating inner cylinder.    
The position resolution of $< 0.1$~pixel yields a resolution of particle   
velocities and fluctuations of better than $0.1$~pixel/frame.   
The upper limit for measurable velocities is given by the tracking   
routine, which requires that the maximum displacement be smaller   
than one particle diameter.  Larger particle images yield a larger velocity   
range, but the velocity profile may not improve since fewer particles   
can be tracked in a single image.  A mean particle size of   
about $20$~pixels gives sufficient dynamic range for the velocities with   
good statistics.      
   
\section{Experimental Results}   
\label{expsec2}   
   
\subsection{Particle Dynamics}   
   
The behavior of the inner cylinder is found to be very similar to    
the dynamics of a rough plate sliding across a granular    
layer~\cite{Nasuno97}.     
At low shear rates the motion of the inner cylinder    
is intermittent with short, rapid slips,    
and long periods of sticking.  At sufficiently high shear rates    
or with a stiff connection between motor and cylinder,    
steady motion of the inner cylinder is observed.     
Air flow reduces the shear forces, as already noted   
by Tardos {\it et al.}~\cite{Tardos98}, and it also suppresses stick-slip    
motion.     
    
\begin{figure}[t]   
\begin{center}   
\epsfig{file=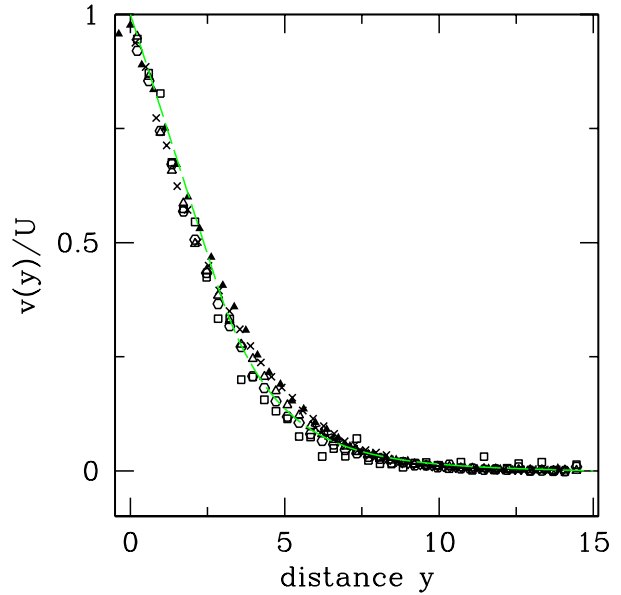,width=8.0cm,height=8.0cm}   
\end{center}   
\caption{Mean particle velocity (normalized by the shear rate)   
 as a function of distance from the inner    
cylinder (in particle diameters) for glass spheres.    
The respective shear rates $U$ (in Hz) are : 0.004 (hexagons), 0.04   
(squares), 0.01 (open triangles), 0.4 (crosses). The   
solid triangles show    
the velocity profile at $U=0.01$ Hz {\it without} air flow.   
The normalized mean velocity profile is independent of   
shear rate and shear dynamics (intermittent or steady motion).   
The dashed line is the   
solution of (\ref{V_profile}), with $\delta=4.7 d$, $y_w=2.8 d$, and   
$\alpha=0.4$ ( see text for details).   
}   
\label{vel_profile}   
\end{figure}   
   
The velocity profile $V(y)$,    
when normalized by the shear rate, is roughly independent of    
shear rate, as shown in Fig.~\ref{vel_profile}.  The mean velocity   
profile without air flow in the stick-slip regime (solid triangles)    
is essentially the same as that for steady shearing.     
The dashed line shows the theoretical prediction from our hydrodynamic   
model, which will be discussed in section~\ref{theorychapter}.   
 
\begin{figure}[t]   
\begin{center}   
\epsfig{file=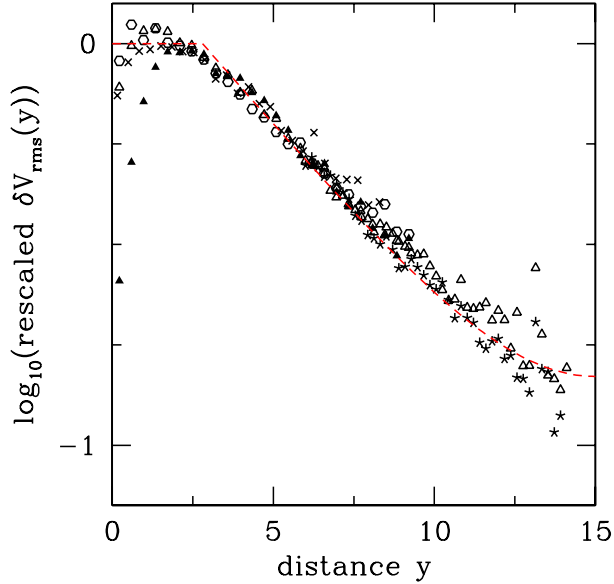,width=8.0cm,height=8.0cm}   
\end{center}   
\caption{RMS velocity fluctuations perpendicular to the shear direction.   
Fluctuations decrease roughly exponentially far from the inner cylinder,   
but more slowly than the mean flow. The rms fluctuations are rescaled    
(shifted vertically) such   
that all experimental points are forced to agree at $y=3 d$, 
{where $d$ is the bead diameter.}  
The dashed line is the theoretical   
result (see text), with a decay length $\delta=4.7 d$ and a boundary position   
$y_w=2.8 d$.  Measurements are made on glass spheres, with symbols as in Fig. 2.}   
\label{vel_fluct}   
\end{figure}

Fig.~\ref{vel_fluct} shows the perpendicular RMS velocity fluctuations   
$\delta V_y$, which   
have not been previously measured in a 3D system to our knowledge.   
When data taken at different shear rates $U$ are   
normalized to the same magnitude at a distance of 3 particle diameters   
away from the wall,   
the fluctuations follow the same profile, independent of shear rate and    
independent of the presence or absence of stick-slip motion.    
The velocity fluctuations decrease roughly exponentially far    
from the inner cylinder but fall off more slowly with $y$ than does    
the average velocity.   
  
We determine fluctuations by averaging the velocity of   
a small section of the image over a long time and then calculating   
deviations of individual particle velocities from the mean of that section.   
The measured parallel fluctuation amplitude therefore includes    
fluctuations of the flow speed, which    
can be caused by the soft spring connection to the motor.    
In order to compare velocity fluctuations during stick-slip motion   
and steady shearing and to compare the measurements to the 
hydrodynamic model of    
a steady state flow, we show only the perpendicular fluctuations.     
We note that even    
during steady shearing, parallel fluctuations are larger than    
perpendicular fluctuations, but their ratio remains roughly independent of    
$y$, as shown in Fig. 5 below.    
   
In principle the density profile could also be measured using the tracking  
algorithm, by counting the average number of tracked particles as a function  
of position $y$. In pratice however no quantitative results could be obtained  
because of the limitation of the tracking method to resolve particle positions  
in the third dimension (e.g., for low densities, particles from lower layers  
are also counted: the density is thus quantitatively overestimated, while 
this does not affect the mean and RMS velocity profiles).  
Qualitatively, the measured density profile increases with the radial 
coordinate toward a limiting value at large distances (not shown). 
The density close to the moving boundary  
is measured to be up to $40$\% below its limiting value, depending on 
shear rate and airflow.  
This is in agreement with other  
numerical and experimental observations \cite{Mueth2000,Schoellmann99}.  
  
We now examine the particle behavior near the boundaries in more detail.    
Close to the   
inner cylinder it is difficult to distinguish wall particles from particles   
that move close to the wall. Because we image the surface  
from a slight angle    
and because the height of particles fluctuates slightly, the boundary   
between wall particles and sheared particles fluctuates.   
We have examined the boundary conditions with both   
steel spheres and rough ceramic   
spheres, which allow us to distinguish the particles from the    
layer of rough glass beads glued to the inner cylinder.   
We find that   
the granular temperature has an approximately constant value in a region   
about $3$ particle diameters wide near the inner wall (see Fig. 3).   
   
\begin{figure}   
 \begin{center}   
\epsfig{file=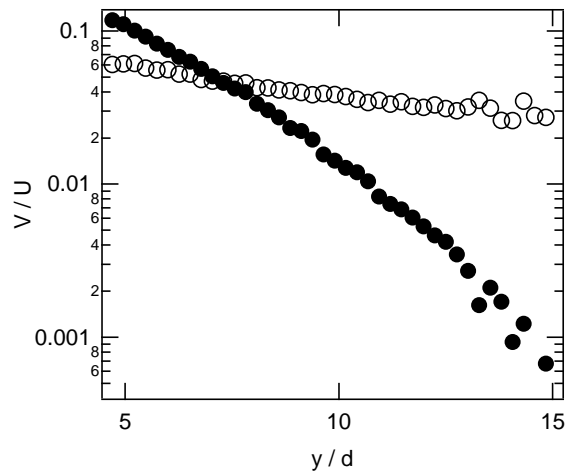,width=8.0cm}   
\end{center}   
\caption{Velocity(solid circles) and fluctuation profile (open circles)  
close to the stationary outer cylinder.    
Each profile is normalized by the shear rate.}   
\label{outer-rim}   
\end{figure}

Since particles barely move close to the stationary outer cylinder,   
the granular temperature at large $y$ is   
examined at a lower frame rate than is necessary  
{at positions} close to the rapidly moving    
inner cylinder.  This yields better   
statistics for fluctuations in particle motion.   
Fig.~\ref{outer-rim} shows the mean velocity    
and velocity fluctuations close to the outer wall.   
The velocity decreases roughly exponentially up to roughly two particle    
diameters from the outer wall.  The velocity fluctuations also decrease   
approximately exponentially in this section.  
   
\begin{figure}[t]   
 \begin{center}   
\epsfig{file=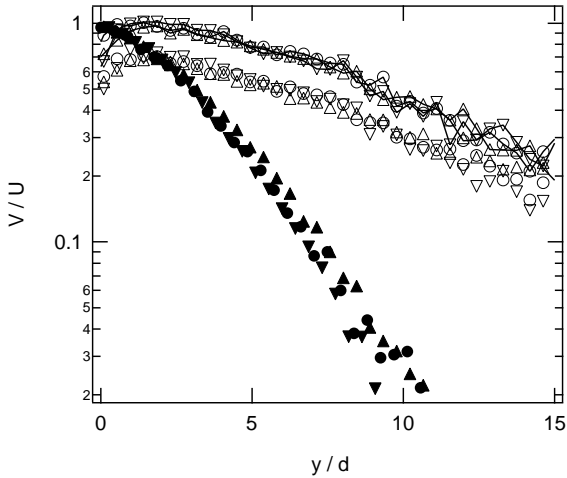,width=8.0cm}   
\end{center}   
\caption{Velocity profile $V(y)$ (solid symbols) and RMS fluctuation profile   
$\delta V(y)$    
(open symbols) both perpendicular (no lines) and parallel (lines)    
to the shear direction   
at different effective pressures controlled by air flow.    
Measurements on glass spheres with upward airflow (triangles pointing up),  
no airflow (circles), and downward airflow (triangles pointing down).   
The profiles do not depend significantly on the pressure.}   
\label{pressure-dep}   
\end{figure}

As discussed in the previous section, an upward    
airflow reduces the effective pressure, while a downward airflow increases   
the effective pressure within the material.  This allows us to   
measure the    
pressure dependence of the velocity and fluctuation profile.   
Fig.~\ref{pressure-dep} shows    
three experiments at different air flow rates (i.e. pressure).   
In order to avoid stick-slip motion without airflow and with downward   
airflow, the motor is connected rigidly to the cylinder for these   
experiments.  The crossing points of  
the temperature and velocity profiles in Figs. 4,5 are different  
because the mean velocity was different.     
   
Neither the velocity profile nor the profile of RMS fluctuations    
(the granular temperature) change 
with pressure over the range of pressures accessible with this method.   
This is consistent with our hydrodynamic model as described in    
section~\ref{theorychapter}.   
Note that the    
RMS fluctuations parallel to the shear direction are larger by a factor of    
roughly $1.3$, even though the mean velocity cannot   
fluctuate due to the rigid connection between motor and cylinder.   
This anisotropy has been observed previously~\cite{campbell86}.   
It may be connected to an anisotropy in pressure in a sheared granular    
system~\cite{jenkins88,goldhirsch96}.    
   
\begin{figure}[t]   
 \begin{center}   
\epsfig{file=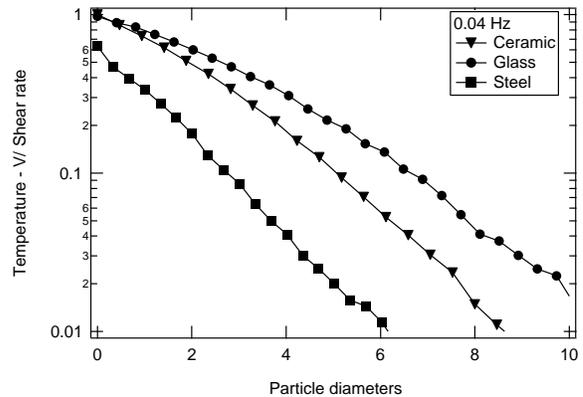,width=8.0cm}   
\end{center}   
\caption{Comparison of $V(y)$ for smooth   
glass spheres (circles), rough porous ceramic particles (triangles),   
and steel sphere mixtures (squares) at $U=0.04 {\rm Hz}$. }   
\label{comparison}   
\end{figure}

The velocity and fluctuation profiles are independent of shear rate   
and shear dynamics, but can vary with the material    
that is sheared.  Our model contains three parameters that determine   
the velocity and fluctuation profiles, as discussed in    
section~\ref{theorychapter}.    
These parameters may depend on material properties in a  non-trivial way.     
In Fig.~\ref{comparison} we    
compare three very different materials (steel spheres, glass spheres,   
rough porous ceramic particles).   
For all materials, the mean velocity decays roughly exponentially far    
from the shear boundary.   
The characteristic length of that decay   
is between $1.5$ and $2$ particle diameters.   
The velocity of glass spheres decreases more slowly away from the  
sheared cylinder than does that of rough ceramic particles.  
Steel spheres are about twice as large as    
the glass spheres coating the surface.  This leads to significant slip    
at the boundary.   The  
RMS fluctuations decay significantly more slowly than the mean velocity    
for all materials (not shown).     
   
\begin{figure}[t]   
 \begin{center}   
\epsfig{file=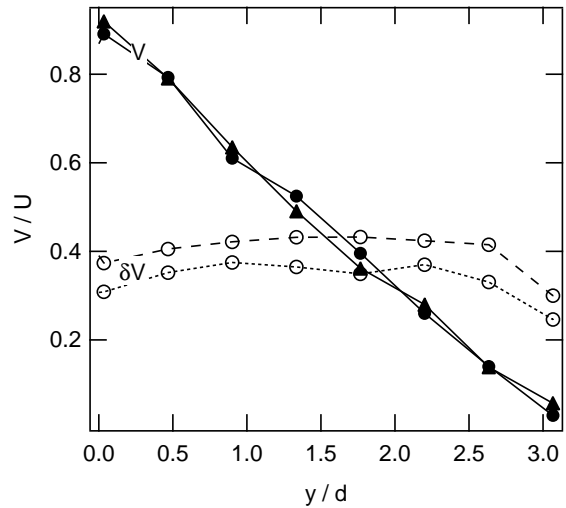,width=8.0cm}   
\end{center}   
\caption{Velocity profiles (Solid lines) and RMS fluctuation profiles 
(dashed lines)  in a narrow gap geometry   
with gap width of $4-5$~particle diameters. The    
velocity profile is linear and the RMS fluctuation profile is constant.  
Measurements are made on glass spheres with $U=0.1 {\rm Hz}$ (circles) and  
$U=0.02 {\rm Hz}$ (triangles).}   
\label{smallgap}   
\end{figure}

We have also done experiments in a narrow gap geometry, where    
the shear region is only $4-5$ particle diameters thick.   
In this case the velocity   
profile is linear and the temperature is roughly constant across the    
cell, as shown in figure~\ref{smallgap}.     
Fluctuations parallel to the   
shear are somewhat larger than perpendicular fluctuations.   
The velocity profile is again independent of shear rate.   
The small gap result is consistent with the hydrodynamic model   
of section~\ref{theorychapter}:   
The uniform RMS fluctuations across the gap are accompanied by a linear   
velocity gradient.

\subsection{Shear forces}   
   
\begin{figure}[t]   
\begin{center}   
\epsfig{file=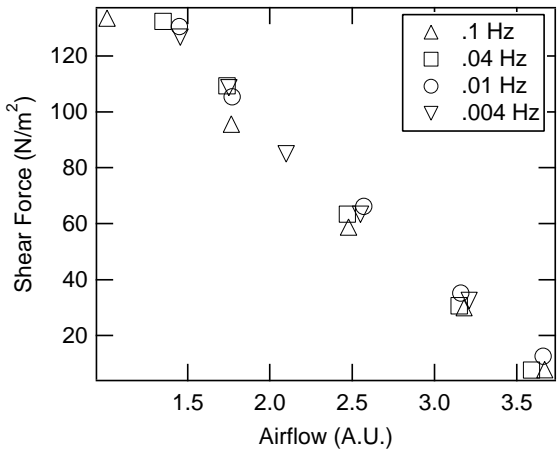,width=8.0cm}   
\end{center}   
\caption{Mean shear stress vs air flow rate at shear rates of 0.004 Hz,   
0.01 Hz, 0.04 Hz, and 0.1 Hz. The shear stress decreases    
approximately linearly with increasing air flow.  The shear stress is    
independent of shear rate at most air flows. (Glass spheres)}   
\label{shearstress}   
\end{figure}

The shear stress is found to be roughly independent of shear    
rate but to decrease roughly linearly with increasing upward   
air flow as shown in figure~\ref{shearstress}.  The dependence on   
air flow is   
consistent with the results by Tardos {\it et al.}~\cite{Tardos98}    
described in section~\ref{backgroundchapter}.   
Previous experiments~\cite{Geminard1999} showed that the shear stress   
is directly proportional to the pressure inside the granular material.   
We can therefore assume that air flow decreases the pressure   
roughly linearly with increasing flow rate.

{When there is no air flow, we also find that the shear stress is 
roughly independent of shear rate even though stick-slip motion is 
observed.  This indicates that some velocity weakening (i.e., a decrease in 
shear force with increasing velocity)} must occur. 
The mean shear stress with air flow is a factor of four smaller than the mean   
shear stress without air flow.     
For a small gap,     
the shear force increases with decreasing shear rate,   
eventually leading to jamming of the inner cylinder below a threshold   
shear rate.     
   
\begin{figure}[t]   
\begin{center}   
\epsfig{file=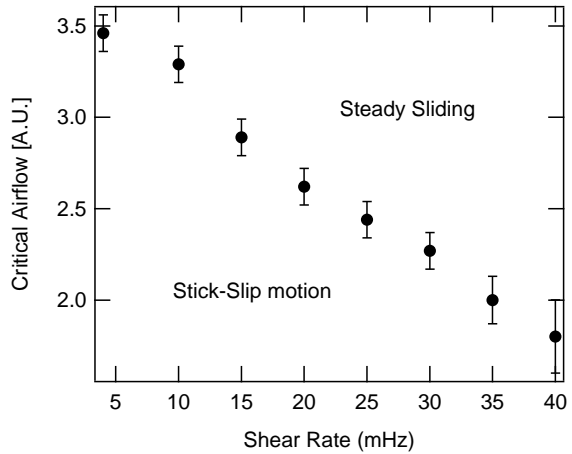,width=8.0cm}   
\end{center}   
\caption{Air flow at which the transition from stick-slip    
motion to steady shearing is observed as a function of the shear rate.   
The critical air flow decreases roughly linearly with shear rate. (Glass spheres)}   
\label{critical_shear}   
\end{figure}

The air flow at which the transition from stick-slip motion   
to steady shearing is observed is shown as a function   
of shear rate in Fig.~\ref{critical_shear}.   
We determine the transition from stick-slip motion to steady sliding motion   
from the emergence of a peak at $v=0$ in the probability    
distribution of the shear velocity.    
The critical air flow decreases roughly linearly    
with shear rate.   
   
\section{Theory}   
\label{theorychapter}   
   
\subsection{Hydrodynamic Model and main assumptions}   
   
Here, we propose a hydrodynamic model for granular   
flow to describe the data presented in the previous section.    
By hydrodynamic, we mean that a {\it local}, mean relationship is assumed   
to hold between shear stress and shear rate, in contrast to    
recent approaches {which advocate nonlocal relationships} 
 (see e.g., \cite{Mills99},    
and \cite{clement99} and references therein).    
The grains are assumed to behave   
like inelastic hard spheres, with a diameter $d$ and    
coefficient of restitution $e$. We therefore {\it neglect} any {\it friction}    
force between    
grains. Within this model, collisions between grains are instantaneous.    
The inelasticity   
coefficient $e$ is moreover assumed to be independent of the relative velocity   
of the two colliding particles. Our philosophy is different from    
more sophisticated approaches (such as the simulations of    
Ref. \cite{Schoellmann99}), where    
the microscopic model is taken to be as    
realistic as possible. Here we deliberately choose the  simplest   
model: our goal is to show that despite its simplicity, a hydrodynamic model    
leads to many non-trivial results which are usually attributed   
in the literature to more sophisticated ingredients of granular flow.   
We check its validity by   
comparing {\it a posteriori} our findings to the experimental results.    
As we demonstrate below, most of the experimental properties of    
sheared granular flow can be explained by this hydrodynamic,    
locally Newtonian description of the material.    
   
We consider a simplified Couette geometry: the granular material is  
confined between two parallel walls, separated by a distance $H$.   
The $x$ axis is along the walls, while the $y$ axis is perpendicular to it.  
The bottom wall, placed at $x=0$, is assumed to move at a velocity $U$  
along the $x$ direction and the top wall stays at rest.  
  
We  start with the equations of hydrodynamics for the inelastic   
hard sphere fluid.   
{Following common practice},  we     
identify the granular temperature with $m [\delta V]^2$,  
where $\delta V$   
is one component ($x$ or $y$) of the RMS velocity fluctuation previously  
defined in    
section~\ref{expreschapter}, and $m$ is the particle mass.  There is a small ambiguity in this definition    
for a sheared granular material since   
the $x$ and $y$ components of $[\delta V]^2$  ({\it i.e.}   
the components parallel and perpendicular to the flow) differ slightly.   
However,   
as may be observed    
in Fig.~\ref{pressure-dep}, they differ only by a scaling factor   
close to unity. We shall therefore disregard this anisotropy in  
our theoretical treatment.   
For the inelastic hard sphere fluid of local particle density $\rho$,    
the equations of hydrodynamics for    
the mean velocity field $V$ and temperature $T$ can be    
written as~\cite{jenkins83}   
\begin{mathletters} 
\begin{eqnarray}   
m &\rho &\left( {\partial V \over \partial t} + V\cdot \nabla V \right)   
= - \nabla \cdot \sigma   
\label{hydroa} \\ 
m &\rho &\left( {\partial T \over \partial t} + V\cdot \nabla T \right)   
= -\nabla \cdot {\bf Q} - {\bf \sigma} : {\bf \kappa} - \epsilon T,   
\label{hydrob}   
\end{eqnarray}   
\label{hydro} 
\end{mathletters} 
where $:$ means contraction of the two tensors.   
In these equations, $\kappa$ is the symmetrized {velocity}-gradient tensor   
\begin{equation}   
\kappa_{\alpha,\beta}= {1\over 2} \left( \partial_\alpha V_\beta   
+ \partial_\beta V_\alpha\right),   
\label{kappa}   
\end{equation}   
${\bf \sigma}$ is the pressure tensor, ${\bf Q}$ is the heat flux,    
and $\epsilon$ is   
the temperature loss rate per unit volume. As in a Newtonian fluid,   
we assume a {\it linear}, {\it local} relationship between fluxes and forces.   
We thus write the pressure tensor as   
\begin{equation}   
{\bf \sigma} = P {\bf I} - 2 \eta \left( \kappa - \nabla \cdot    
V {\bf I} \right),   
\label{sigma}   
\end{equation}   
where $P$ is the pressure, $\eta$ is the shear viscosity and ${\bf I}$ the    
unit tensor.   
In a similar way, we assume Fourier's law for the heat flux,   
\begin{equation}   
{\bf Q}= - \lambda \nabla T,   
\label{fourier}   
\end{equation}   
with $\lambda$ the thermal conductivity.   
These equations are completed by the equation of state of the material,    
in the form $P=P(\rho,T)$.   
   
In the simplified planar shear cell geometry, the mean flow   
is a function of $y$ only and parallel to the $x$ direction: ${\bf V}=   
V(y) {\bf e}_x$ with ${\bf e}_x$ the unit vector in the $x$ direction.    
>From the momentum transport equation (\ref{hydroa}) {in the steady 
state} one then expects   
\begin{eqnarray}   
& &{\partial \over \partial y} \sigma_{yx} = 0 \\   
& &{\partial \over \partial y} \sigma_{yy} = 0 .    
\label{conserv_momentum}   
\end{eqnarray}   
This shows that both the shear stress $\sigma_{xy}$ (=$\sigma_{yx}$)   
and the pressure $P=\sigma_{yy}$ are independent of $y$.   
   
\subsection{Equation of state and high-density expressions for the    
transport coefficients}   
   
In order to solve the hydrodynamic equations (\ref{hydro}),    
explicit expressions   
for the transport coefficients and the equation of state in terms of    
density and temperature are needed.    
  
When the system is at rest, the density $\rho$ in the granular material is    
roughly given by the random close   
packing (RCP) density $\rho_c$ (the granular system does not to 
crystallize  because of    
a slight polydispersity of the beads). Experimentally  
{shear rate}   is found to dilate the  system~\cite{Bagnold1954}.  
More precisely the density decrease is larger close    
to the boundaries, where the  {shear rate} 
is larger, than far from the moving   wall,    
where the velocity goes to zero \cite{Behringer99,Schoellmann99}. 
As already quoted in Sec. III., our experimental results for the density 
do qualitatively agree with this observation.   
Consequently {\it a broad range of densities} is explored in the sheared    
system, going from a value  
slightly below the RCP density  far from the moving wall, down to 
a density up to $40\%$  less than RCP 
near the moving wall at high shear rates.    
This makes the problem much more difficult than in standard fluids,   
where the density remains constant over the cell. Here the functional    
dependence   
of the transport coefficients on density is an important ingredient of   
the theory since it does affect flow properties.   
   
\subsubsection{Equation of state}   
   
For an inelastic hard sphere system, the equation of state can be    
written in terms   
of $g(d)$ the pair correlation function at contact     
($d$ being the diameter) \cite{JPH,Dufty1999},   
in the form   
\begin{equation}   
P=\rho T~\left[1+(1+e){\pi \over 3} \rho~d^3~g(d)\right].   
\label{EOS}   
\end{equation}   
($\rho$ is the local numerical density, {\it i.e.} the number of  
particles per unit volume).  
For both dilute and moderately dense systems ($\rho d^3 \sim 1$),    
$g(d)$ is accurately described   
by the Carnahan-Starling formula \cite{JPH}. However since the density of the granular   
material in the shear cell is close to random close packing (RCP) where   
$g(d)$ diverges,   
an alternative expression for $g(d)$ is usually assumed  
~\cite{RCP}:   
\begin{equation}   
g(d)= {1 \over {1-{\rho/\rho_c}}},   
\label{g_d}   
\end{equation}   
with $\rho_c$ the density at RCP.   
The equation of state, Eqs.\ (\ref{EOS})-(\ref{g_d}),    
then takes the following approximate    
form in the high density limit   
\begin{equation}   
P=\rho_0 ~{1 \over (1-{\rho\over \rho_c})}~T,   
\label{EOS2}   
\end{equation}   
with $\rho_0=(1+e){\pi \over 3} \rho_c^2 d^3$. The equation of state thus gives   
a natural link between temperature and density: since the pressure is independent of $y$, one obtains   
\begin{equation}   
1-{\rho\over \rho_c}= {\rho_0 \over P}~T.   
\label{rhoT}   
\end{equation}   
This equation shows that the regions of small temperature correspond to    
high density regions, $\rho \sim \rho_c$, and vice-versa.   
   
\subsubsection{Transport coefficients}   
   
Expressions for the transport coefficients of the inelastic    
hard-sphere model have been   
computed from the Enskog    
equation~\cite{jenkins83,Dufty1999}.   
This kinetic equation takes the effects of excluded volume   
into account, but it neglects any correlation    
between the velocities of colliding particles.    
This approach yields expressions for the transport coefficients in   
terms of the density,   
temperature, and pair correlation function at contact.   
We refer to Refs.~\cite{jenkins83,Dufty1999} for explicit expressions.   
We note that {\it in the high density limit}, $g(d)$ becomes   
very large and the dependence of the Enskog transport coefficients on    
density mainly come from the terms proportional to $g(d)$. In this case,   
the transport coefficients reduce to the generic forms   
\begin{eqnarray}   
\eta_E(\rho,T) &\simeq &\eta_0 {m^{1/2} \over d^2}~g(d)~T^{1/2} \nonumber\\   
\lambda_E(\rho,T) &\simeq& \lambda_0 {1\over {m^{1/2}  d^2} }~g(d)~T^{1/2}    
\nonumber\\   
\epsilon_E (\rho,T) &\simeq& \epsilon_0  
\rho {1\over {m^{1/2}  d}}~g(d)~T^{1/2},   
\label{Enskog}   
\end{eqnarray}   
where  $m$ is the mass of the particles and    
$\eta_0$, $\lambda_0$, $\epsilon_0$ are dimensionless   
coefficients that depend only slightly on density in the high-density limit.   
If $\rho\sim \rho_c$ as discussed above, these coefficients can be taken as   
numerical constants. Finally, we mention that $\epsilon_0$ is proportional to   
$1-e^2$, where $e$ is the coefficient of restitution.  Thus $\epsilon_0=0$ in the    
purely elastic case, as expected.   
   
The full expressions for the transport coefficients    
obtained within the Enskog kinetic theory    
have been found {in simulations} to be correct for small and    
intermediate densities~\cite{Alder1970}.    
However, as mentioned above, the density of the flowing material    
in the shear cell is higher and close to the RCP density.   
In this limit, some of the Enskog expressions for the transport coefficients   
may no longer be valid, mainly because correlations between colliding particles   
and collective phenomena, which are not included in the Enskog theory,    
then play an important role. The reason is quite   
intuitive: at high density close to random close packing, a particle can move    
substantially (over a distance of the order of its diameter) only if    
its neighbors move coherently. Only collective motion is therefore possible.   
It has been found in Molecular Dynamics simulations of the hard sphere    
model~\cite{Alder1970}   
that these correlations only affect the shear viscosity and self    
diffusion coefficient,   
which depart from their Enskog approximation at high densities. On the   
other hand, the thermal conductivity has been found to be well described  
by the Enskog  expression up to very high densities.   
Such behavior is in fact expected since transport of energy   
does not require motion of particles over large distances    
(only ``rattling" around the   
mean position is involved in energy transport). Similar conclusions   
were reached by Leutheusser~\cite{Leutheusser1982}     
on the basis of a mode-coupling calculation for the elastic hard-sphere model.   
   
As a result of these considerations,   
the thermal conductivity $\lambda$  and loss rate $\epsilon$ are expected    
to keep their Enskog expressions $\lambda_E$ and  $\epsilon_E$, as given    
in Eqs.~(\ref{Enskog}),   
over the whole range of densities (between intermediate densities up to RCP):   
\begin{eqnarray}   
&\lambda(\rho,T) &\simeq \lambda_0 {1\over {m^{1/2}  d^2}}~{1 \over    
\left(1-{\rho\over \rho_c}\right)}~T^{1/2} \nonumber\\   
&\epsilon (\rho,T) &\simeq \epsilon_0 {1\over {m^{1/2}  d}}~{\rho_c     
\over \left(1-{\rho\over \rho_c}\right)}~T^{1/2}.   
\label{transport_0}   
\end{eqnarray}   
On the other hand, a crossover is expected for the shear viscosity between    
its Enskog approximation, in Eq.~(\ref{Enskog}), for intermediate densities,   
towards an asymptotic stronger divergence as a function of density very    
close to   
RCP. Such a crossover is indeed found in Molecular Dynamics calculations of   
the self diffusion coefficient in a monodisperse hard sphere system close to    
freezing~\cite{Woodcock81}.   
By analogy with the behavior of    
supercooled liquids above the glass transition~\cite{gotze},    
we shall assume that    
very close to RCP, the viscosity diverges    
{\it algebraically} as a function of density $\rho$ near $\rho_c$:   
\begin{equation}   
\eta(\rho,T) \sim {\eta_1\over \left(1-{\rho\over \rho_c }\right)^\beta}   
 {m^{1/2} \over d^2}~T^{1/2} .   
\label{visc}   
\end{equation}   
Here $\eta_1$ is   
a dimensionless numerical constant.   
At this point the exponent $\beta$ is a phenomenological    
parameter in the theory.   
Since the viscosity is expected to diverge more strongly    
than its Enskog expression, we   
expect $\beta$ to be larger than one.   
We shall discuss in the next section how the form proposed in 
Eq.~(\ref{visc}) compares with the experimental results.   
  
We emphasize again that such an algebraic divergence of the viscosity is  
 expected in supercooled liquids close to the glass transition \cite{gotze}. More precisely, an algebraic divergence is associated with  
the existence of cooperative interactions which predominate over  
thermally activated processes \cite{gotze}.  The functional 
form in Eq.\ (\ref{visc}) is predicted by mode-coupling calculations 
of the dynamics of supercooled liquids. 
In the case of 3D hard sphere, the latter approach yields an exponent 
$\beta=2.58$ \cite{Barrat89}. 
 
To our knowledge, there is no expression for the viscosity over a    
full range of densities that would make    
explicit the cross-over between the Enskog   
expression at intermediate densities and the asymptotic expression at   
very high densities.    
In order to avoid restrictive assumptions about the general   
form of the density dependence of the viscosity,    
we shall write in general   
\begin{equation}   
\eta(\rho,T) = \bar{\eta}(\rho) {m^{1/2} \over d^2}~T^{1/2} ,    
\label{visc_0}   
\end{equation}   
where the function $\bar{\eta}(\rho)$ has the two limiting forms:   
\begin{eqnarray}   
&\rho \sim d^{-3}< \rho_c\qquad & \bar{\eta}(\rho)   
\simeq {\eta_0 \over \left(1-   
{\rho\over \rho_c }\right)} \nonumber \\   
&\rho \sim \rho_c \qquad& \bar{\eta}(\rho) \sim {\eta_1\over \left(1-   
{\rho\over \rho_c }\right)^\beta} .   
\label{eta}   
\end{eqnarray}   
As we shall show in the following sections, the knowledge of    
these two limiting    
behaviors is sufficient to obtain a quantitative description of the flow and a    
qualitative picture for the shear forces.   
   
Combining Eqs.\ (\ref{EOS2}) and (\ref{transport_0}) allows us to write   
the transport coefficients    
in terms of the pressure and temperature  as   
\begin{eqnarray}   
&\eta(\rho,T) &\simeq {m^{1/2} \over d^2}~\bar{\eta}\left({\rho_0 T \over P}\right)~T^{1/2} \\   
&\lambda(\rho,T) &\simeq \lambda_0 {1\over {m^{1/2}  d^2}}~{P \over \rho_0 T^{1/2}}~ \\   
&\epsilon (\rho,T) &\simeq \epsilon_0 {1\over {m^{1/2}  d}}~  {P \over  T^{1/2}}.~   
\label{transport}   
\end{eqnarray}   
(Note that in order to improve readability, we dropped a numerical constant prefactor   
$\rho_c/\rho_0$ in the expression for $\epsilon$, which amounts to a    
rescaling of the numerical prefactor $\epsilon_0$.)   
   
\subsection{RMS and Mean flow velocity profiles}   
   
In this section we compute the granular temperature profile, $T(y)$ and   
mean-velocity profile, $V(y)$. In the stationary   
Couette geometry, the hydrodynamic equation for $T(y)$ is found to reduce to   
\begin{equation}   
{\partial \over {\partial y}} \lambda(\rho,T){\partial \over {\partial y}} T   
+\sigma_{xy} \dot \gamma - \epsilon (\rho,T) T =0,   
\label{T_equ}   
\end{equation}   
and   
\begin{equation}   
\sigma_{xy} = \eta(\rho,T) \dot\gamma ={\rm const},   
\label{cst_sigma}   
\end{equation}   
where $\dot\gamma={d V_x(y)\over dy}$ is the shear rate.   
Using the expression for the shear stress,    
$\sigma_{xy}=\eta(\rho,T) {\dot \gamma}$,    
the second term of Eq.\ (\ref{T_equ}) can be rewritten as   
$\sigma_{xy} \dot \gamma=\sigma_{xy}^2/\eta(\rho,T)$.    
{Note that $\sigma_{xy}$ is negative because ${\dot \gamma} = \partial v_y 
/\partial x $ is negative throughout the sample.} 
   
Both equations involve the explicit form of the density dependence of the   
viscosity, $\eta(\rho,T)$, for which no explicit functional form   
is available to us. At first sight, it would seem    
hopeless to obtain a full expression of the temperature and velocity profiles.    
This is not the case. As we shall show, a simple    
phenomenological picture, which emerges from the asymptotic forms   
of $\eta(\rho,T)$ as given in Eq.\ (\ref{eta}), allows one to    
overcome this problem and to obtain tractable    
expressions for the temperature and mean velocity.   
   
We proceed in two steps. First, we describe this   
phenomenological ``two region'' picture and obtain expressions   
for the velocity and temperature profile. Then   
in a second step we come back to a more general but formal solution   
of Eqs.\ (\ref{T_equ}) and (\ref{cst_sigma}) (in the next section).    
This general discussion allows us to discuss the velocity    
dependences of the shear forces.   
   
We start with the discussion of the temperature profile. The role of the    
the nonlinear term $\sigma_{xy} \dot \gamma$ in Eq.\ (\ref{T_equ})   
is in fact physically quite simple to understand. It merely   
acts as a source (``heating'') term for the fluctuations:    
it is through this nonlinear contribution   
that the flow creates the fluctuations which {\it in fine} couple back to   
the mean flow. However, this source term is only effective   
close to the wall as we show below. This simplifies considerably the   
picture for the creation and transport of temperature. Basically,   
two regions can be treated separately: close to the moving boundary,   
fluctuations are ``created'' through the nonlinear coupling to the flow;   
far from the boundaries, inelastic dissipation dominates over nonlinear  
heating, and the local temperature profile is determined entirely by the  
diffusion and heat loss terms of Eq.\ (\ref{T_equ}). 
   
This separation can be understood   
on the basis of the asymptotic behavior of the viscosity discussed above.    
Indeed, it is easy to show that far from the wall    
the nonlinear term $\sigma_{xy} \dot \gamma$   
goes to zero faster than the dissipation term $\epsilon (\rho,T) T$    
in Eq.~(\ref{T_equ}).   
Away from the moving boundary the temperature goes to zero and the density    
goes accordingly towards RCP. From Eqs.~(\ref{eta})    
and~(\ref{transport}) it follows that the nonlinear    
term behaves in this region like     
${\sigma_{xy}^2/ \eta(\rho,T)} \propto T^{(2\beta-1)/2}$, while the dissipative   
term in this region scale with temperature as $\epsilon(\rho,T) T    
\propto T^{1/2}$   
(where Eq.~(\ref{transport}) has been used).    
Since we anticipated that the exponent   
$\beta$ is larger than one (around $1.75$ as found experimentally, see below),   
the nonlinear term decays more strongly towards zero than the dissipative one.    
On the other hand, the nonlinear    
term is relevant close to the moving boundary where the   
shear rate is large (or equivalently the density is smaller). We note that   
if the exponent $\beta$ had been equal to $1$, both previous terms   
would have been comparable and then the picture of two separate    
regions would not have been appropriate.   
   
Close to the walls, the full nonlinear equation including   
the ``heating'' term should be solved. This is in fact    
{not necessary} to make   
simple predictions for   
the RMS and mean velocity profiles. An inspection of the experimental   
profiles, as presented in   
Fig.~\ref{vel_fluct}, shows that in the region close to   
the moving boundary,  the temperature is roughly constant over a   
small layer, several particle    
diameters thick, with a thickness which can be considered   
in a first approximation to be independent of the shearing velocity $U$.    
As a first step, we therefore assume    
in an ad hoc way that for distances $y$ smaller than a cut-off distance $y_w$,   
the temperature $T$ is constant, $T(y)=T_0$.     
In this pictorial view,   
the boundary layer corresponds to the region where the nonlinear    
term is important.   
At this stage, the parameter $T_0$ merely    
plays the role of a normalizing constant.   
We shall come back to this point in much more detail in the discussion of   
the shear forces because, while the precise value of $T_0$ does not   
influence the velocity profile,  it does strongly affect   
the prediction for the shear forces.   
   
We are now in a position to obtain an explicit expression for the temperature   
profile.   
For distances larger than $y_w$, the transport equation for the   
temperature, Eq.~(\ref{T_equ}), reduces to    
\begin{equation}   
{\partial \over {\partial y}} \left\{\lambda_0 {1\over {m^{1/2}  d^2}}~   
{P \over \rho_0 T^{1/2}} {\partial \over {\partial y}} \right\}  T   
 - \epsilon_0 {1\over {m^{1/2}  d}}~  P  T^{1/2} =0,   
\label{T_equ1}   
\end{equation}   
where the high-density expressions for the transport coefficients,   
Eqs.~(\ref{transport}), has been used.   
   
As shown previously, the pressure $P$ is    
independent of $y$, so one may rewrite   
Eq.~(\ref{T_equ1}) as   
\begin{equation}   
{\partial^2 \over {\partial y^2}}   T^{1/2} -  {1\over \delta^2} T^{1/2}=0,   
\label{T_equ_final}   
\end{equation}   
where $\delta$ has the dimension of a length and is defined as   
\begin{equation}   
\delta^2={2 \lambda_0\over{\epsilon_0 \rho_0~d}}.   
\label{delta}   
\end{equation}   
The parameters $\lambda_0$, and $\epsilon_0$   
are just numerical constants in the high-density regime of interest to us.   
Therefore, one expects $\delta$ to be of the order   
of a few particles diameters.  Note that since $\epsilon_0$    
is proportional   
to $1-e^2$  
(with $e$ the restitution coefficient), the decay length   
$\delta$ goes to infinity when the system become perfectly elastic,    
as one would expect.   
This equation has to be completed by boundary conditions    
for the temperature at both walls.    
At the moving wall, we set $T(y)=T_0$ for $y<y_w$, as discussed above.   
At the wall at rest, detailed experiments, as shown in Fig. \ref{outer-rim},   
show that the temperature profile   
is compatible with a vanishing heat flux condition   
$dT/dy=0$. (In general, one expects the   
boundary condition for the temperature to relate the heat    
flux at the boundary to  the product of  
the interface (Kapitza) resistance and the temperature jump:    
Here we just assume   
that the Kapitza resistance is very small. We shall come back    
to this point in the next section.)   
   
The solution of Eq.~(\ref{T_equ_final}) with these boundary conditions is   
\begin{eqnarray}   
&y<y_w \qquad& T(y)=T_0 \\   
&y_w<y \qquad& T^{1/2}(y) = T_0^{1/2} {\cosh\left({ H-y\over \delta}\right) \over   
\cosh\left({ H-y_w\over \delta}\right)  }.   
\label{RMS_profile}   
\end{eqnarray}   
In this equation, $H$ is the thickness of the shear cell.    
As shown in Fig.~\ref{vel_fluct}, this result is in good agreement with    
the experimental data. We find that  $\delta=4.7 d$ and   
$y_w=2.8 d$ for glass spheres, but these parameters depend somewhat on the   
material properties.   
   
\begin{figure}[] \begin{center}   
\epsfig{file=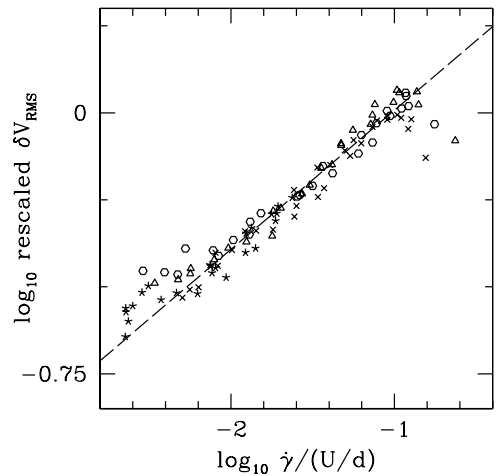,width=6.5cm}   
\end{center}   
 \caption{Connection between the local RMS velocity fluctuations and local   
shear rate (same symbols as for Fig.~\ref{vel_profile}).    
Local fluctuations are found to increase approximately as a power law    
of the local velocity gradient, with a power of $0.4$ (dashed line). }   
\label{fluct_grad}   
\end{figure}

The mean velocity profile can be obtained along the same lines from   
the temperature profile of Eq.\ (\ref{RMS_profile}).     
Eq.~(\ref{cst_sigma})  gives unambiguously the shear rate    
$\dot\gamma$ in terms of the temperature profile $T(y)$,    
since the $y$ dependence of the density is already   
contained in the temperature through the equation of state,    
Eq.~(\ref{rhoT}).    
It seems at first sight difficult to make an explicit prediction    
for the velocity profile,    
since the explicit expression of the viscosity is not known over the full range   
of densities.  However, the ``two regions'' picture, which emerged in the   
discussion of the temperature profile, is relevant for the mean-velocity   
profile as well.   
Far from the wall ({\it i.e.} for $y>y_w$),   
the temperature decays to zero and   
the density goes accordingly to the RCP limit    
(as shown in Eq.~(\ref{rhoT})). In this region,   
one thus expects an ``anomalous'' scaling for the density dependence of   
the viscosity, according to Eq.\ (\ref{eta}). Thus one finds for $y>y_w$   
\begin{equation}   
\eta_0 {m^{1/2} \over d^2}~\left({P \over \rho_0 T}\right)^\beta~T^{1/2}  
~ \dot\gamma = \sigma_{xy}.   
\label{scal_0}   
\end{equation}   
It is interesting to note that   
this relation yields a scaling relationship between the temperature    
$T$ and $\dot\gamma$    
that can be tested experimentally. The previous equation can be rewritten   
in the form   
\begin{equation}   
\dot\gamma = \left\{ {\sigma_{xy} \over {\eta_0 {m^{1/2} \over d^2}~ 
\left({P \over \rho_0}\right)^\beta}} \right\}    
~T^{ {1\over 2}(2\beta-1)}.   
\label{scaling_theo}   
\end{equation}   
This power law relationship has been checked experimentally,    
as shown in Fig.~\ref{fluct_grad},   
where the local RMS velocity profile $\delta V \equiv (T/m)^{1/2}$    
is plotted versus the local velocity gradient $dV_x/dy$ on logarithmic scales.    
   
In this plot a scaling exponent $\alpha=0.4$ is obtained    
for the glass bead system. This allows for an   
experimental determination of the exponent $\beta$ for the ``anomalous''    
divergence of   
the viscosity close to RCP. According to Eq.\ (\ref{scaling_theo}),   
$\alpha$  is identified in the theory as  $\alpha=1/(2\beta-1)$,    
so that $\beta=1.75$. The fact that this exponent clearly exceeds unity   
confirms the collective character of the dynamics in the granular material. 
This value of the exponent is slightly smaller than the previously mentioned 
result for $\beta=2.58$ obtained within the mode coupling theory  
applied to the case of hard spheres \cite{Barrat89}.  
A possible reason for this  
difference could lie in the rotational degrees of freedom of the beads 
in the experimental system which are absent from the mode  
coupling estimate. By increasing the number of degrees of freedom  
an effective decrease of the shear viscosity could be obtained.  
We leave this question for further investigation.  
 
We emphasize that the observation of the scaling relationship 
between $\dot\gamma$ and $T^{1/2}$   
is not restricted to our system. We have applied the  
same procedure to the data of Howell {\it et al.} \cite{Howell99}, which  
were obtained in a 2D system of photoelastic disks : for the 14 different   
densities studied in this work, a similar scaling relationship   
between the local $\delta V_{RMS}$ and $\dot\gamma$ is measured.  
An exponent  $\alpha \sim 0.5$ is obtained, yielding $\beta \sim 1.5$ for  
that experiment.  
Moreover  both the mean and fluctuation profiles obtained  
by Howell {\it et al.} can be fitted within the present model.  
This observation seems to indicate that the coupling between mean and  
RMS fluctuations which is the essence of our model is a general feature  
of the underlying dynamics of the granular material.

Once the exponent $\beta$ is determined,    
the velocity profile can be obtained by integration.   
In the boundary layer for the temperature ($y<y_w$), the temperature is   
assumed to be a constant; Eq.~(\ref{cst_sigma}) then shows that the   
shear rate is constant, yielding a linear velocity profile.    
This observation is in    
agreement with the experimental velocity profiles, as can be seen by a   
careful inspection of Fig.~\ref{vel_profile} for $y<y_w$.   
On the other hand for $y>y_w$ the velocity profile is obtained by   
integrating Eq.~(\ref{scaling_theo}) together   
with the solution (\ref{RMS_profile}) for the temperature profile. The full   
velocity profile is obtained by matching the two solutions at $y_w$ (where we   
assume continuity of velocity and shear rate).  The solution   
obeying no-slip boundary conditions at both walls takes the form   
\begin{equation}   
V_x(y)= U \left(1 - {\int_0^y dy^\prime~\phi(y^\prime) \over {\int_0^H dy^\prime~\phi(y^\prime)}}\right)   
\label{V_profile}   
\end{equation}   
where the function $\phi(y)$ is defined as   
\begin{eqnarray}   
& y<y_w\qquad & \phi(y) = 1 \\   
& y>y_w\qquad & \phi(y)= \left[{\cosh\left({ H-y\over \delta}\right) \over   
\cosh\left({ H-y_w\over \delta}\right)  }\right]^{2\beta-1}.   
\label{phi}   
\end{eqnarray}   
   
When the value $\beta=1.75$ previously obtained for the exponent $\beta$ is   
used, the theoretical velocity profile   
defined in Eqs.\ (\ref{V_profile},\ref{phi}) is easily integrated    
numerically. The result is plotted in Fig.~\ref{vel_profile}    
together with the experimental results.    
As can be seen, good agreement is obtained with the theory.   
   
Although the two region picture seems to be quite    
successful in describing the mean and RMS velocity profiles, it would be more    
satisfactory to have a full expression of the density dependence of the    
viscosity over the full range of densities in order to integrate    
Eqs.~(\ref{T_equ}) and~(\ref{cst_sigma}) explicitly.    
The two region picture is to be considered merely as a simple and    
physically sound way of dealing with our ignorance of the explicit   
functional dependence of the viscosity on density. Its success indicates    
that the velocity profile does not depend crucially on the details of this    
relationship.

\subsection{Towards ``Solid like'' Shear Forces}   
   
\subsubsection{General discussion}   
   
In his pioneering work, Bagnold measured the shear force in a    
Couette cell and found a quadratic increase of the shear force    
as a function of the shearing velocity, $\sigma_{xy} \propto U^2$    
\cite{Bagnold1954}). Using kinetic arguments he proposed a   
 phenomenological relationship between shear stress   
and shear rate in the form $\sigma_{xy} \propto \dot\gamma^2$,  which can be    
understood by assuming that the collision frequency between granular particles    
is fixed by the shear rate $\dot\gamma$ itself. Subsequent experiments   
were performed \cite{Tardos98} which showed that the functional    
dependence of the shear force as a function of the shearing velocity   
does depend on whether the system is allowed to dilate or not:   
when the system is sheared at constant volume, the shear force is   
found to be proportional to $U^2$ ({\it i.e.} the Bagnold scaling); when   
the system is allowed to dilate, the shear force is found    
to be independent of $U$.   
In the present geometry, the material is allowed to dilate and a solid-like    
friction force,i.e. approximate independence of $U$, is found.   
   
It is tempting to ascribe the solid like behavior to a solid    
friction force existing between  granular particles at contact.     
On the other hand it is difficult to reconcile this behavior with    
constant-volume Bagnold scaling, $\sigma_{xy} \propto U^2$, 
whose origin is purely kinetic.  
Here we shall show that both behaviors can be accounted for within the    
hydrodynamic model we have introduced.    
   
\begin{itemize}   
\item{Constant volume results: }    
   
It is evident that the Bagnold scaling, $\sigma_{xy} \propto   
U^2$, originates from a general dimensional argument,    
whatever the local relationship between shear stress and shear rate is:   
in other words, the {\it global} scaling $\sigma_{xy} \propto U^2$ obtained   
in a constant volume experiment does not necessary imply that    
the {\it local} relationship $\sigma_{xy} \propto \dot\gamma^2$.    
The reason is very simple. Let's consider a system of $N$ inelastic    
hard spheres contained in a plane Couette cell of (fixed) volume $V$.   
The bottom boundary is moving at a constant    
velocity $U$, while the upper boundary stays at rest.    
   
In this case, the only microscopic velocity scale entering the dynamics    
is $U$, the velocity of the   
moving wall. In other words, {\it all dynamical quantities} can be written   
in terms of the shear   
velocity $U$ only, which fixes the only time scale in the problem.    
>From the analysis of the full $N$-body dynamics, one thus concludes in general 
that     
\begin{eqnarray}   
{ V_x(y)\over U} & = &f\left({y\over d},{N\over V}, {H\over d}\right) 
\nonumber\\   
{T(y)\over{m U^2}} & = &g\left({y\over d}, {N\over V}, {H\over d}\right)  
\nonumber\\   
P& =& {N\over V} p\left({N\over V},{H\over d}\right)~m U^2 \nonumber\\   
\sigma_{xy} & = & {N\over V} \sigma_0\left({N\over V},{H\over d}\right)~m U^2 , \\   
\label{Bagnold_scaling}   
\end{eqnarray}   
where $f$, $g$, $p$, $\sigma_0$ are dimensionless functions; $H$ is the cell    
width and $d$ the bead diameter.    
   
In this context, the Bagnold scaling results naturally from the last equality    
in~(\ref{Bagnold_scaling}), {\it whatever the relationship between   
$\sigma_{xy}$ and $\dot\gamma$ is}. The only important condition is   
that the microscopic dynamics of the granular material only involve   
binary collisions.    
   
\item{Constant pressure results:}    
   
Experiments performed in a constant pressure situation   
are more subtle to handle.  The introduction   
of a constant pressure in the system introduces a new time scale,   
namely $(m/Pd)^{1/2}$, which will compete with the time scale    
associated with the shearing velocity, $d/U$. Thus different regimes   
might be found depending on whether the time scale associated with the   
shear rate is larger or smaller than the one associated with pressure.   
\end{itemize}   
   
We show in the following   
that in the situation of constant pressure, a shear force independent of    
shear rate is obtained at large shearing velocity, while a velocity weakening   
regime is found at small velocities.   

\subsubsection{Shear forces at constant pressure from the hydrodynamic model: qualitative}   
   
In the previous section the experimental mean and RMS   
velocity profiles were reproduced   
by using a simple `` two regions'' picture, which allowed us to overcome our   
lack of knowledge of the full functional dependence of the viscosity on the   
density.    
However the shear force  originates in the interaction between the   
wall and the fluid particles adjacent to it. Therefore one needs a good   
description    
of the region close to the wall in order to get valuable information about   
the shear forces. This requires a careful analysis of the nonlinear   
``heating'' term in the equation for the temperature, Eq.~(\ref{T_equ}),   
which we perform in this section.   
   
We first present an approximate, qualitative argument that helps  
explain physically what quantities control the friction force    
in the shearing   
experiment. We then present the full approach, which clarifies   
the crude approximations made in the qualitative arguments.  
   
The shear force is defined in the hydrodynamics model as the viscosity   
times the shear rate, both evaluated {\it at the boundary}. Using the expression   
for the viscosity, Eq.~(\ref{visc_0}), this yields:   
\begin{equation}   
\sigma_{xy}= \bar{\eta}\left( {\rho_0 T_0 \over P} \right) {m^{1/2} \over d^2} T_0^{1/2} \dot\gamma\vert_0.   
\label{shear0}   
\end{equation}   
We need a second equation giving the temperature at the boundary   
$T_0$.    
In the ``two regions'' picture we introduced above, the temperature   
is created near the boundary and then transported to the rest of the system   
In this picture, the temperature $T_0$ merely comes from a balance    
between the dissipation rate close to the moving boundary $-\epsilon T_0$   
and the ``heating term'', $\sigma_{xy} \dot\gamma\vert_0$.     
Together this gives   
\begin{equation}   
\sigma_{xy}\dot\gamma\vert_0 \simeq {\epsilon_0 \over m^{1/2} d} P T_0^{1/2},   
\label{heat0}   
\end{equation}   
where we have used the expression in Eq.~(\ref{transport})    
for the dissipation rate $\epsilon$.   
   
The shear rate $\dot\gamma\vert_0$ can be eliminated from the two previous   
equations, Eqs.~(\ref{shear0}) and~(\ref{heat0}), to give    
\begin{equation}   
{\sigma_{xy} \over P}  \simeq \left({ \epsilon_0 \over \rho_0d^3}   
\bar{\eta}\left( {\rho_0 T_0 \over P} \right) {\rho_0 T_0 \over P}    
\right)^{1/2}.   
\label{sig_P}   
\end{equation}   
   
Now, depending on the ratio $\rho_0 T_0 / P$, two limiting behaviors are   
obtained\\   
\indent (i) If $\rho_0 T_0/ P$ is large, then the viscosity takes its   
intermediate density expression in Eq.\ (\ref{eta}),    
$\bar{\eta}(\rho_0 T_0/ P) \simeq \eta_0 P/\rho_0T_0$. In this limit,   
one then obtains from Eq.\ (\ref{sig_P})   
\begin{equation}   
{\sigma_{xy} }  \simeq \mu_0 P   
\label{friction_solide_0}   
\end{equation}   
with $\mu_0={ \epsilon_0 \eta_0 /\rho_0d^3}$ a dimensionless constant.   
In other words, the shear force is expected to be independent of the    
shear rate and proportional to pressure, as is usually found in    
solid friction.\\   
\indent (ii) If $\rho_0 T_0/ P$ is small, then the viscosity takes its   
high density expression in Eq.\ (\ref{eta}),   
$\bar{\eta}(\rho_0 T_0/ P) \simeq \eta_1 (P/\rho_0T_0)^\beta$ and    
Eq.\ (\ref{sig_P}) gives   
\begin{equation}   
{\sigma_{xy} \over P}  \simeq \left({ \epsilon_0 \eta_1\over \rho_0d^3}   
\left( {P \over \rho_0 T_0 } \right)^{\beta-1}    
\right)^{1/2}.   
\label{int_0}   
\end{equation}   
The temperature $T_0$ enters explicitly in this formula. It has to be obtained   
as a function of the shearing velocity $U$. This can be accomplished using   
the energy balance discussed above, which results in Eq.\ (\ref{heat0}).   
In this equation, one expects the shear rate at the boundary to be   
roughly given by $\dot\gamma\vert_0 \sim U/\ell_0$, where $\ell_0$    
is a distance typically of the order of a few diameters    
(this point is confirmed in the full discussion to follow). One thus gets    
\begin{equation}   
T_0^{1/2} \sim {\sigma_{xy} \over P} {d \over \ell_0} m^{1/2}U,   
\label{temp_0}   
\end{equation}   
which still depends on the ratio between shear stress $\sigma_{xy}$   
and pressure $P$.   
Combining this expression for $T_0$ as a function of $U$ and $\sigma_{xy}/   
P$ and the formula for $\sigma_{xy}/P$ as a function of $T_0$ in   
Eq.\ (\ref{int_0}) yields the final result   
\begin{equation}   
{\sigma_{xy} } \simeq \mu_1 P \left( {\rho_0 mU^2 \over P} \right)^{1-\beta \over {2\beta}},   
\label{WV_0}   
\end{equation}   
where we have introduced the dimensionless constant, $\mu_1=   
\left({\epsilon_0 \eta_1\over \rho_0d^3} ({d \over \ell_0}^{1-\beta})^{1-\beta}   
\right)^{1\over {2\beta}}$.   
   
The cross-over between the two regimes occurs at the ``critical''   
velocity $U_c$ defined as $U_c\sim(P/m\rho_0)^{1/2}$.   
   
\subsubsection{Shear forces: quantitative}   
  
These results obtained from the arguments of Sec. V.D.2 are in fact fully    
confirmed by a more detailed and careful analysis of the nonlinear set of    
equations    
for  temperature and velocity, Eq.\ (\ref{T_equ}) and Eq.\ (\ref{cst_sigma}).   
Here we present the full analysis.   
In order to simplify the discussion    
we assume that the granular material   
is semi-infinite, {\i.e.} the cell width $H$    
is larger than the decay length for   
temperature $\delta$ defined in Eq.\ (\ref{delta}).   
   
We proceed in two steps. First we obtain two closed implicit equations    
for the temperature at the boundary ($T_0$) and the shear stress $\sigma$   
from an analysis of Eqs.\ (\ref{T_equ}) and (\ref{cst_sigma}). These equations   
are written in terms of the general form of the viscosity, Eq.\ (\ref{visc_0}).   
Then in a second step, we discuss the different limiting behavior for   
the shear stress as a function of the shear velocity    
$U$ (``large'' and ``small'' velocities).   
   
\paragraph{\bf First condition defining $T_0$ and $\sigma_{xy}$ : }   

We first rewrite the full temperature equation, Eq.\ (\ref{T_equ}), in terms of the   
expressions for the transport coefficients,  Eqs.\ (\ref{visc_0}) and     
(\ref{transport_0}) :    
\begin{eqnarray}   
{\partial \over {\partial y}} \left\{ {\lambda_0 \over {m^{1/2}  d^2}}~   
{P \over \rho_0 T^{1/2}} {\partial \over {\partial y}} \right\}  T   
+ {\sigma_{xy}^2 \over {\bar{\eta}(\rho) {m^{1/2} \over d^2}~T^{1/2} }}   
&&\nonumber \\   
- \epsilon_0 {1\over {m^{1/2}  d}}~  P  T^{1/2} & = &0,   
\label{T_equ_full}   
\end{eqnarray}   
which can be rewritten as    
\begin{equation}   
{\partial^2 \over {\partial y^2}}   T^{1/2}     
+ {T_s \over {\delta^2\bar{\eta}(\rho)T^{1/2}}}   
-  {1\over \delta^2} T^{1/2}=0,   
\label{T_equ_full1}   
\end{equation}   
where we have introduced    
$T_s=\delta^2\sigma_{xy}^2 \rho_0 d^4/(2\lambda_0 P)$, a quantity with   
the dimension of a temperature; It can be rewritten more conveniently    
as $T_s=({P/\rho_0}) \left({\sigma_{xy}\over P}\right)^2~   
{\rho_0d^3 \over \epsilon_0}$,   
if we use the explicit expression of $\delta$   
which has been given in Eq.\ (\ref{delta}).   
It is important to note that the term $\bar{\eta}(\rho(y))$ is   
a function of $T(y)$ only   
through the equation of state, Eq.\ (\ref{EOS2}).   
Let us introduce $u=(T/m)^{1/2}$, which has the dimension of a velocity.   
A first integral of Eq.\ (\ref{T_equ_full}) can be obtained using   
a standard recipe   
of classical mechanics: we multiply Eq.\ (\ref{T_equ_full}) by $du/dy$ and   
integrate over $u$. We obtain:   
\begin{equation}   
{1\over 2} \left( {du\over dy} \right)^2 +   
\int_0^u du^\prime {T_s \over m \delta^2}   
{1 \over \bar{\eta}(\rho_0 m u^{\prime 2}/P) u^\prime}   
-{1\over 2 \delta^2} u^2=E.   
\label{first_integral}   
\end{equation}   
Note that the density dependence    
of $\bar{\eta}(\rho)$ has been rewritten in terms   
of its dependence on $u=(T/m)^{1/2}$ thanks to the equation of state,   
Eq.\ (\ref{rhoT}).   
The parameter $E$ is a constant which in the   
mechanical analogy fixes the ``energy''   
of the system.   
We now introduce the effective ``potential'' $V(u)$ defined as   
\begin{equation}   
V(u)= \int_0^u du^\prime {T_s \over m \delta^2}   
{1 \over \bar{\eta}(\rho_0 m u^{\prime 2}/P) u^\prime} -{1\over 2 \delta^2} u^2 ,   
\label{potential}   
\end{equation}   
which allows us to rewrite Eq.\ (\ref{first_integral}) as   
\begin{equation}   
{1\over 2} \left( {du\over dy} \right)^2 + V(u)= E.   
\end{equation}   
The behavior of the potential $V(u)$ can be obtained by analyzing   
the two limits $u\rightarrow 0$ and $u\rightarrow \infty$.   
The limit $u\sim 0$ corresponds to small temperature, {\it i.e.} densities   
close to RCP.   
According to the expression of the viscosity close to RCP, Eq.\ (\ref{eta}),   
the first term in the potential in    
Eq.\ (\ref{potential}) behaves in this limit like    
\begin{eqnarray}   
& & \int_0^u du^\prime {T_s \over m \delta^2}   
{1 \over \bar{\eta}(\rho_0 m u^{\prime 2}/P) u^\prime}   
\nonumber \\  
& & \simeq {T_s d^2 \over {\eta_1 m{3/2}\delta^2}}   
\left( {\rho_0 m \over P }\right)^{\beta}   
\int_0^u du^\prime u^{\prime 2\beta -1}   
 \propto u^{2\beta}.   
\label{behav1}   
\end{eqnarray}   
This term {vanishes more rapidly than the last,    
quadratic term of $V(u)$   
(since the exponent verifies $\beta >1$). Therefore close to $u\sim 0$, one   
has $V(u) \simeq -{1\over 2 \delta^2} u^2$.    
   
In the limit of large $u$, {\it i.e.}, high temperature and intermediate   
density, the integral term in $V(u)$ behaves as   
\begin{eqnarray}   
\int_0^u du^\prime~{T_s \over m \delta^2}   
{1 \over \bar{\eta}(\rho_0 m u^{\prime 2}/P) u^\prime}   
&\simeq& \int_0^u du^\prime~{\rho_0 T_s \over {\eta_0 P \delta^2}}  u^{\prime}  \nonumber \\   
&=&{\rho_0 T_s \over {\eta_0 P }} {u^2 \over {2 \delta^2}},   
\label{behav2}   
\end{eqnarray}   
which, when combined with the second term in Eq.\ (\ref{potential}), gives    
\begin{equation}   
V(u)=\left[{\rho_0 T_s \over {\eta_0 P }} -1\right] {u^2 \over {2 \delta^2}}.   
\label{pot2}   
\end{equation}   
Thus, depending on the sign of the first factor on the right had side of  
the equation, $V(u)$ goes to plus or minus infinity   
when $u\rightarrow \infty$. In Fig. \ref{Potential},  
we have arbitrarily chosen a positive sign   
to emphasize the different behavior of $V(u)$ in the different limits,  
but the discussion does not depend on this particular choice.   
\begin{figure}[t]    
\begin{center}   
\epsfig{file=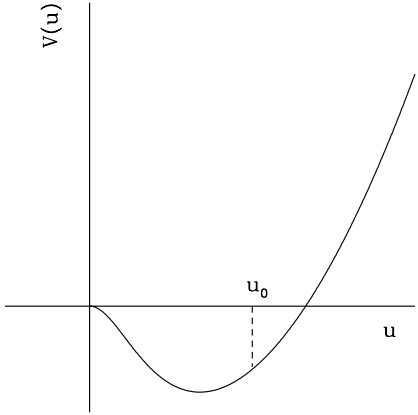,width=6.5cm,height=6.5cm}   
\end{center}   
 \caption{Typical behavior of the potential $V(u)$ as a function of $u$.}   
\label{Potential}   
\end{figure}   
   
The ``energy'' $E$ in Eq.\ (\ref{first_integral}) can be computed from  
the boundary   
conditions. As already mentioned, we assumed a semi-infinite system in this   
section to simplify the discussion. At infinity,  
both the temperature and its gradient   
are expected to vanish, and so does $u$ : $du/dy\rightarrow 0 $  
and $u\rightarrow 0$.   
Consequently the ``energy'' $E$ is obtained to be zero : $E=0$.   
This imposes the temperature at the boundary, $T_0=m u_0^2$,    
which has to satisfy   
\begin{equation}   
{1\over 2} \left. \left( {du\over dy} \right)^2\right\vert_{y=0} + V(u_0) =0.   
\label{T0_1}   
\end{equation}   
As we mentioned briefly in Sec. V.C,    
one expects in general that the boundary condition for the temperature    
relates the temperature gradient to the temperature at the    
boundary, in the form   
\begin{equation}   
J_Q=-\kappa \left. {dT\over dy}\right\vert_{y=0} = R_K T\vert_{y=0},   
\label{kapitza}   
\end{equation}   
where $R_K$ is a phenomenological parameter, usually    
denoted as the Kapitza resistance.   
The ratio $\ell_K=\kappa/R_K$ has the dimension of a length. Following the same   
arguments which led us to the expression of the  
thermal conductivity, Eq.\ (\ref{transport}),   
$R_K$ is expected to be proportional    
to the fluid-wall collision frequency, {\it i.e.}   
to $P/(\rho_0 T^{1/2}(y=0))$. The length $\ell_K$ is thus expected to be independent of    
pressure, and only fixed by the ``microscopic'' quantities like the diameter of the particles   
and the roughness of the walls. Rewriting Eq.\ (\ref{kapitza})  
in terms of the field $u$, we find the condition for the temperature at the 
boundary     
\begin{equation}   
{1\over 2 \ell^2} u_0^2 + V(u_0) =0,   
\label{T0_BC}   
\end{equation}   
with $\ell=2\ell_K$. Using Eq.\ (\ref{potential}) for $V(u)$, we obtain    
\begin{equation}   
{1\over 2}\left( {1\over \ell^2}-{1\over \delta^2 } \right)u_0^2   
+   
\int_0^{u_0} du {T_s \over m \delta^2}   
{1 \over \bar{\eta}(\rho_0 m u^{2}/P) u} = 0,   
\label{T0_BC_final}   
\end{equation}   
where we recall that $T_0=mu_0^2$ and  $T_s   
=({P/\rho_0}) \left({\sigma_{xy}\over P}\right)^2~   
{\rho_0d^3 \over \epsilon_0}$.   
   
It is interesting to note that,   
since the second term in Eq.\ (\ref{T0_BC_final})   
is positive, the first term has to be negative, so that $\ell>\delta$.   
If the opposite case, the only allowed solution to Eq.\ (\ref{T0_BC_final})   
is $T_0=0$, {\it i.e.} no temperature is introduced in the system.   
This condition is in some sense expected : if $\ell$ goes to zero,   
the boundary condition for the temperature at the wall, Eq.\ (\ref{kapitza}),   
imposes $T_0=0$, so that no temperature can be introduced in the system.   
Physically $\ell$ is related the amount of energy input at the boundary,   
while $\delta$ has to do with the dissipation (see Eq.\ (\ref{delta})),   
so  the condition $\ell >\delta$ can be understood as an energy balance   
condition : if not enough energy is introduced in the system at the boundary,   
the dissipation is too strong to allow flow.   
   
\paragraph{\bf Second condition defining $T_0$ and $\sigma_{xy}$ : }   

A second condition relating $\sigma_{xy}$ and $T_0$ can be obtained by   
formally integrating the momentum equation, Eq.\ (\ref{cst_sigma}):   
\begin{equation}   
V_x(y)=U-\int_0^y {|\sigma_{xy}| \over \eta (\rho,T)} dy^\prime ,   
\end{equation}   
{where we used the fact that $\sigma_{xy} <0$ to replace $\sigma_{xy}$ 
by $- |\sigma_{xy}|$}   
As $y\rightarrow \infty$, the velocity goes to zero, so  one gets the   
condition  
\begin{equation}   
U=\int_0^\infty {|\sigma_{xy} |\over \eta (\rho,T)} dy^\prime.   
\label{cond2}   
\end{equation}   
The explicit form of the $y$ dependence of density and temperature is   
not known explicitly, but can be obtained formally from   
the first integral obtained previously,  
Eq.\ (\ref{first_integral}) (with $E=0$). The latter    
yields an implicit condition for the field $u$ as   
\begin{equation}   
dy= - {du \over {\sqrt{-2V(u)}}},   
\label{EOM}   
\end{equation}   
which allows us to rewrite the second condition, Eq.\ (\ref{cond2}),    
in terms of $\eta(\rho,T)$ only   
\begin{equation}   
U={|\sigma_{xy}| d^2 \over m}  
\int_0^{u_0}{1 \over \bar{\eta}(\rho_0 m u^{ 2}/P) u}   
{du \over \sqrt{-2V(u)}},   
\label{SIG_BC_final}   
\end{equation}   
where the potential $V(u)$ is given in Eq.\ (\ref{potential}).   
   
The two equations, Eqs.\ (\ref{T0_BC_final}) and  (\ref{SIG_BC_final}), in   
principle allow both $T_0$ and $\sigma_{xy}$ to be determined in terms of the   
pressure $P$ and velocity $U$.   
   
\paragraph{\bf Limiting behaviors of the shear forces}   

Two limits of the previous equations can be discussed for large    
and small shear velocity limit $U$ (we shall specify below what we   
mean by ``large'' and ``small'').   
   
\indent{\bf (i) Large velocity limit}   
   
Let us assume that the temperature at the boundary $T_0$ satisfies   
$T_0 > P/\rho_0$. This corresponds to the large velocity limit.   
In this case, the density close to the wall is intermediate and   
the shear viscosity takes its Enskog asymptotic expression, as    
quoted in Eqs.\ (\ref{visc_0}) and (\ref{eta}) :    
$\bar{\eta}(\rho_0 m u^{ 2}/P)\simeq P/\rho_0 m u^{ 2}$.   
   
In Eq.\ (\ref{T0_BC_final}), corresponding to the first condition,    
the integral is dominated by its   
large $u$ behavior, where the function $\bar{\eta}$ behaves like   
$\bar{\eta}(\rho_0 m u^{2}/P)\simeq (P/m\rho_0) u^{-2}$.   
To a good  approximation the first condition then takes the asymptotic form   
\begin{equation}   
{1\over 2}\left( {1\over \ell^2}-{1\over \delta^2 } \right)u_0^2    
+   
\int_0^{u_0} du {T_s \over  \delta^2}   
{\rho_0  u \over {\eta_0 P} }=0,   
\end{equation}   
which can be rearranged to give   
\begin{equation}   
\left( {1\over \ell^2}-{1\over \delta^2 } +{\rho_0 T_s \over {\delta^2\eta_0 P}}    
\right)u_0^2=0.   
\end{equation}   
Using the expression for $T_s$, one gets    
\begin{equation}   
\sigma_{xy} = \mu_d P,   
\label{solid_friction}   
\end{equation}   
with $\mu_d=\left\{{\eta_0 \epsilon_0 / \rho_0d^3}(1-{\delta^2\over   
\ell^2})\right\}^{1/2}$.   
   
Although the model we have developed is purely a hydrodynamic model, this    
relationship is usually referred as to {\bf solid-like} behavior.   
   
The temperature at the boundary can be obtained from the second condition,   
Eq.\ (\ref{SIG_BC_final}). The integral on the right hand side of   
Eq.\ (\ref{SIG_BC_final}) is dominated by the behavior near $u\simeq u_0$,   
with $\bar{\eta}(\rho_0 m u^{ 2}/P)\simeq P/\rho_0 m u^{ 2}$ and   
$V(u) \simeq V(u_0)={-u_0^2/2\ell^2}$. One thus gets   
\begin{eqnarray}   
&U&\simeq {\sigma_{xy} d^2 \over m^{1/2}} {\ell \over u_0}   
\int_0^{u_0} du~{m\rho_0 u \over {\eta_0 P}} \nonumber \\   
& &={\sigma_{xy} \over P}  m^{1/2}   
{{\rho_0 d^2\ell} \over {2\eta_0}} u_0,   
\end{eqnarray}   
which gives eventually   
\begin{equation}   
T_0=m u_0^2= \zeta_d~mU^2,   
\label{T0_largeU}   
\end{equation}   
where the numerical prefactor $\zeta_d$ is defined as $\zeta_d=\left({2\eta_0 /({\mu_d \rho_0 d^2 \ell})}\right)^{1/2}$.   
   
This shows that the large velocity regime which we defined as $T_0>P/\rho_0$   
ratifies $U> (P/(\mu_t m\rho_0))^{1/2}\sim (Pd^3/m)$.   
   
\indent{\bf (ii) Small velocity limit}   
   
In this case the viscosity is expected to behave ``anomalously'',   
with $\bar{\eta}(\rho_0 m u^{ 2}/P)\simeq \eta_1 (P/\rho_0 m u^{ 2})^{\beta}$.   
In this limit, the first condition, Eq.\ (\ref{T0_BC_final}), then takes the    
form   
\begin{equation}   
0={1\over 2}\left( {1\over \ell^2}-{1\over \delta^2 } \right)u_0^2   
+   
\int_0^{u_0} {du \over u} {T_s \over  m\eta_1 \delta^2}   
\left({m\rho_0  u^2 \over {P}}\right)^\beta,   
\end{equation}   
which can be reorganized to give   
\begin{equation}   
\sigma_{xy}=P \left[ \left(1-{\delta^2\over \ell^2}\right)   
{\beta \eta_1\epsilon_0 \over \rho_0 d^3}\right]^{1/2}   
\left({\rho_0 m u_0^2 \over {P}}\right)^{{1-\beta}\over 2}.   
\label{a_0}   
\end{equation}   
This equation has to be completed by the second condition, Eq.\   
(\ref{SIG_BC_final}). In this equation, the potential term $V(u)$   
can be approximated by $V(u)\simeq {1\over 2\delta^2} u^2$, since   
the temperature $T=mu^2$ is small (see the discussion of the shape   
of the potential in the previous subsection). Together with the    
asymptotic expression of the viscosity in this limit, one can rewrite   
Eq.\ (\ref{SIG_BC_final}) as   
\begin{equation}   
U=   
{\sigma_{xy} d^2 \over m \eta_1} \int_0^{u_0}   
\left({m\rho_0  \over P}\right)^\beta u^{2\beta-1}   
{du \over \sqrt{{1\over \delta^2} u^2}},   
\end{equation}   
which yields   
\begin{equation}   
U=   
\left({P\over m\rho_0}\right)^{1/2} {\sigma_{xy}\over P} {\rho_0 d^2\delta \over 2\beta-1} \left({\rho_0 m u_0^2 \over {P}}\right)^{{2\beta-1}\over 2}.   
\label{a_1}   
\end{equation}   
   
When combined with Eq.\ (\ref{a_0}), one gets the following relationship    
between shear stress $\sigma_{xy}$ and $U$ :   
\begin{equation}   
\sigma_{xy}= \mu_W P \left\{ {U^2 \left({m\rho_0 \over P}\right)}\right\}^{1-   
\beta\over 2\beta}.   
\label{VW}   
\end{equation}   
The temperature at the boundary is then given as    
\begin{equation}   
T_0=\zeta_W {P \over \rho_0} \left\{{U^2 \left({m\rho_0 \over P}\right)}\right\}^{1\over 2\beta},   
\label{T0_VW}   
\end{equation}   
where the two numerical prefactors $\mu_W$ and $\zeta_W$ are defined   
as $\mu_W=((1-{\delta^2\over \ell^2}) {\epsilon_0 \over \rho_0 d^3}   
\beta\eta_1)^{(2\beta-1)/(2\beta)} ({2\beta-1 \over \rho_0 d^2 \delta   
})^{(1-\beta)/\beta}$ and   
$\zeta_W=((2\beta-1)/(\mu_W \rho_0 d^2\delta))^{1/(2\beta-1)}$.   
{The stress crosses over from its low-velocity power-law in velocity 
behavior to its high-velocity constant value at $U\approx U_c = 
P/(m\rho_0)$.} 
   
A crucial feature of the $U$ dependence of shear stress in this   
regime is that it is {\bf velocity weakening}, {\it i.e.}, the   
shear stress   
{decreases with increasing $U$ for $U< U_c$ because $\beta >1$.}

\subsubsection{General remarks about shear forces at constant pressure}   
   
Before ending this section, we would like to emphasize a few   
points :   
\begin{itemize}   
\item we have identified two regimes in the shearing velocity dependence of the   
shear force. For {\it small velocities}, the shear force is {\it velocity weakening},   
as obtained in Eq.\ (\ref{VW}). For {\it large velocities}, the shear force   
is found to be ``{\it solid-like}'', {\it i.e.}, 
independent of the shearing velocity, and proportional   
to pressure, as shown in Eq.\ (\ref{solid_friction}).   
\item The cross-over between one regime and the other is found   
to occur at a {\it critical velocity}, $U_c=(Pd^3/m)^{1/2}$.   
   
\item  In the velocity weakening regime, {\it i.e.} for velocities   
smaller than $U_c$, the steady sliding regime is expected to be   
unstable. Indeed, as $\partial \sigma_{xy} / \partial U <0$, the   
steady sliding   
situation is unstable to small fluctuations (see Refs. \cite{Elmer,Persson} for   
a full discussion). As a result,    
one might expect a {\it stick slip behavior of the system for $U<U_c$}.   
   
\item Finally, it is interesting to note that   
both the large and small $U$ limits of the shear rates,   
Eqs.\ (\ref{solid_friction}) and (\ref{VW}), are in agreement with our   
previous dimensional analysis of the dynamics in the constant   
volume case (see Sec. V.D.1).   
Since on dimensional grounds the pressure should be proportional   
to $U^2$ when the volume is fixed, one finds that both expressions   
reduce to the Bagnold scaling $\sigma_{xy} \propto U^2$. We emphasize   
that this ``global'' scaling holds even if the local Bagnold   
relationship $ \sigma_{xy} \propto \dot\gamma^2$ is not expected to   
hold in the small velocity regime (where the viscosity exhibits   
an ``anomalous'' behavior).   
    
\end{itemize}

\section{Conclusions}   
\label{discussionchapter}   
   
In this paper, we have investigated  both experimentally and theoretically   
the shear dynamics of  granular flow.   
The granular material was sheared in a Couette cell, instrumented to allow   
measurements of both shear forces and  ``microscopic'' dynamics of   
individual particles.   
The latter measurements were  
performed by tracking the instantaneous positions of   
particles on the upper  
surface using a fast camera and an imaging procedure. A variable  
upward air flow through the granular material allows the  
internal pressure to be adjusted.   
Experimentally the following results were found:   
(i) the flow is restricted to a small region close to the moving boundary;   
(ii) the normalized velocity profile is independent of the shearing   
velocity, pressure (i.e. airflow), and of the type of   
motion of the moving cylinder (stick-slip or continuous sliding);   
(iii) the shear force acting on the moving cylinder is independent of the   
shearing velocity and proportional to pressure, as  shown in the planar geometry  
in~\cite{Geminard1999});   
(iv) there is a close connection between local fluctuations and mean flow,   
as manifested in the   
power law relationship measured experimentally between these two local   
quantities (Fig. 10);   
(v) the RMS velocity profile decays more slowly as a function of the radial  
coordinate than  does  
the mean velocity profile. We  
find that these general features occur for several different types of  
particles, though there are quantitative differences.

On the basis of these observations, we have proposed a ``hydrodynamic''   
model for the granular flow, in which the granular material is assumed    
to behave like a locally Newtonian fluid.  In contrast to ``classical''    
fluids however, the temperature and density of the material are not   
constant over the shear-cell. The temperature, defined in terms of the   
fluctuations of the   
velocity, is created at the moving boundary and propagated through   
the material. The temperature profile thus results from a balance between   
heat flux and local energy loss due to the inelasticity of collisions.    
As a consequence, the density and temperature dependences of the transport   
coefficients play a crucial role.     
  
In the high density regime under consideration, simple,   
asymptotic expressions   
for the transport coefficients can be obtained within the Enskog   
approximation. The latter is   
however expected to be invalid for the density dependence of the viscosity   
at very high densities   
(close to Random Close Packing, RCP) where collective rearrangements    
comes into play. This led us to   
assume a stronger divergence of the viscosity as a function of the density   
close to RCP. In a manner analogous to   
what is usually proposed in supercooled liquids, we have assumed an   
algebraic divergence   
of the viscosity near the RCP density.   
Using the equations of transport of momentum and heat, we then compute the mean and RMS   
velocity profiles. Those are found to be in very good agreement with the   
experimental results. Moreover, the   
scaling law relationship between the mean and RMS velocity profiles that is   
found experimentally is also   
predicted from the model.   
   
Finally, the velocity dependence of the shear force is determined. 
Two regimes are predicted,   
depending on whether the velocity is larger or smaller than a critical   
velocity $U_c=(Pd^3/m)^{1/2}$.   
For large velocities a {\it solid like friction force} is predicted, {\it   
i.e.} independent of the   
shearing velocity, in agreement with the experimental observation. For   
small velocities, a {\it velocity   
weakening} regime is predicted, consistent with the  
occurrence of stick-slip motion in   
this case.   
We emphasize that these results are found within the hydrodynamic model,   
even if no solid friction   
force is assumed to hold between the grains.   
  
In our work, we have assumed that the flow on the upper surface is close  
to that within the interior of the Couette cell, as found in earlier  
work~\cite{Mueth2000}.  However, since the particles on the surface are  
less constrained, their fluctuations could be somewhat different from those  
in the interior, even when we apply a downward pressure through an airflow.  
While  MRI and X-ray measurements in the interior~\cite{Mueth2000} could not 
resolve fluctuations, recent unpublished measurements of fluctuations on the  
more constrained bottom surface layer by the Chicago group~\cite{Jaeger}  
reveal fluctuation profiles similar to ours: Fluctuations decay roughly  
exponentially with a characteristic length longer than that for the mean  
velocity.  The velocity profiles found in these two experiments differ 
somewhat; however differences in geometry (inner and outer radii),  
material properties, and measurement methodology make a direct comparison  
between the experiments difficult.  
  
The main point of the present paper is the agreement between our  
measured velocity and temperature profiles (and friction force) and the  
predictions of a Newtonian hydrodynamic model with a strong density  
dependence of the viscosity taken from models of the glass transition.  
  
Many questions are still open. One important problem, which is not discussed   
here, is the  temporal response   
of the system when a velocity step is imposed. Such an experiment   
would provide information about the transport   
mechanism within the granular material. The response can be predicted   
in principle from the hydrodynamic model and could provide an independent    
test of its validity.   
   
\section{Acknowledgments}   
We thank D. Howell and S. Luding for providing data to us,  and M.  
Ernst, H. Jaeger, J. Jenkins, and S. Nagel  for valuable discussions.  JPG  
acknowledges helpful discussions at the Aspen Center for Physics. This work  
was supported by the National Science Foundation under Grants DMR-9704301  
and DMR-0072203 to Haverford College, and DMR-9730405 and DMR-9632598 to  
The University of Pennsylvania.


\begin{thebibliography}{99}
\bibitem{neglect} We neglect air flow mediated particle-particle interactions.
\end{thebibliography}
\end{document}